\documentclass[amsmath,amssymb,amsbsy,reprint,prb,preprintnumbers,showpacs,superscriptaddress]{revtex4-2}
\usepackage{graphicx,color}
\usepackage{dcolumn}
\usepackage{bm}
\usepackage{braket}
\usepackage{mathtools}
\usepackage{ulem}
\usepackage[breaklinks,colorlinks=true,linkcolor=blue,urlcolor=blue,citecolor=blue]{hyperref}
\usepackage{times}
\usepackage{physics}
\usepackage{latexsym}
\usepackage{amsmath, amssymb}
\usepackage{mathtools}
\usepackage{multirow}

\newcommand{\be}{\begin{eqnarray}}
\newcommand{\ee}{\end{eqnarray}}

\renewcommand{\theequation}{\arabic{equation}}

\begin{document}

\title{Hierarchy of mixed symmetry protected topological states in extended cluster states under subsystem decoherence}
\date{\today}
\author{Yoshihito Kuno} 
\thanks{These authors equally contributed}
\affiliation{Graduate School of Engineering Science, Akita University, Akita 010-8502, Japan}
\author{Takahiro Orito}
\thanks{These authors equally contributed}
\affiliation{Department of Physics, College of Humanities and Sciences, Nihon University, Sakurajosui, Setagaya, Tokyo 156-8550, Japan}

\begin{abstract}
We study the effect of subsystem decoherence to an extended cluster state which is a symmetry protected topological (SPT) phase. The model includes many subsystem $Z_2$ symmetries.
We report that subsystem decoherence induces local charge fluctuations, leading to a mixed SPT state in the unaffected subsystems.
If we start from the extended cluster state, hierarchical mixed-state SPT phases emerge in response to step-by-step subsystem decoherences. These mixed-state SPT phases keep strong symmetries the symmetry of which is protecting symmetries for the initial cluster SPT. Moreover, these SPTs can be characterized by R\'{e}nyi-2 string orders. Then, as the subsystems are progressively decohered, the hierarchy of mixed-state SPT phases terminates in a $Z_2$ strong-to-weak spontaneous symmetry breaking (SWSSB) state on the final remaining subsystem, where a long-range entangled state appears, namely a glassy Greenberger-Horne-Zeilinger (GHZ) state. Our work demonstrates that decoherence is not merely a destructive process, but can induce and organize series of nontrivial mixed-states. This reveals a systematic route from mixed-state SPT order to SWSSB with the glassy GHZ-type long-range entanglement.
\end{abstract}


\maketitle
\section{Introduction} 
Decoherence applied to a pure quantum many-body state transforms it into a mixed state. 
Traditionally, decoherence has often been viewed as detrimental to delicate quantum properties, since it tends to wash away non-trivial properties of pure states or to just lead to featureless mixed ones. 
More recent studies, however, have revealed that decoherence can induce an exotic non-trivially mixed states, having no counterpart in pure states.
In this sense, decoherence can be a key source to generate quantum many-body systems.

Much recently, effects of decoherence to non-trivial many-body pure states well known in condensed matter physics are progressing rapidly \cite{Wen_text,wang2026}. 
These advances have been underpinned by a refined understanding of symmetries in mixed states.
In particular, mixed states have two notions of symmetry, strong or weak \cite{Buca_2012,groot2022}.
This distinction of symmetry helps researchers to explore the mixed state phase of matter with greater detail than ever before.
As one of recent hot topics, application of the decoherence to symmetry-protected topological (SPT) leads to interesting non-trivial mixed states, such as average SPT \cite{ma2023,ma2024,Lee2025,Ma_PRXQuantum6, Guo_2025,Guo2024_2,Guo-and-Ashida2024}. Topological order under decoherences can be changed into intrinsic mixed state topological order where this state cannot be realized as a ground state of a local gapped Hamiltonian \cite{Fan_2024,KOI2024_IMTO,Ellison_PRXQuantum_2025,Sohal2025,Zhang_2025,Wang_2025,kataoka2026,Cai2026}. 
Decoherence applying to density matrix can lead to extensive notion of spontaneous symmetry breaking, namely strong-to-weak spontaneous symmetry breaking (SWSSB) \cite{lee2023,Lessa2024_2,sala2024,KOI2024,chen2024,Guo_PRX_2025,teh2025,Guo_2025,haga2026}.
Remarkably, even though a decoherence channel and its decohered state are invariant to a strong symmetry, the strong symmetry can be spontaneously broken to a weak one, and the symmetry-broken mixed state can simultaneously have a long-range order in the thermodynamic limit.   
Motivated by these advances, considerable effort has been devoted to characterizing and detecting non-trivial mixed states \cite{Lessa2024_2,weinstein2024}.  
Recent experimental observations of SWSSB have further highlighted the importance of understanding in mixed state physics~\cite{Wang_EXP2026}. 

From another perspective, there has also been growing interest in mixed-state physics.
Specifically, applying quantum measurements to a many-body state, along with feedback unitary that depends on the measurement outcome, can generate a specific entangled state in the subsystems of unmeasured sites. 
Recent studies \cite{Raussendorf2005,verresen2022,verresen2022,Tantivasadakarn2022,Lu2023_feedback,KOI_2024_gc} have revealed that various unconventional quantum states can be prepared,
such as ``cat state'' with long-range order (LRO), SPTs, topological ordered states, fractons, and non-Abelian topological ordered states.
The technique used to create the cat state can be extended to general cluster SPT states~\cite{KOI_2024_gc}. 
 
Although these works primarily focus on the realization of mixed states that retain characteristic features of non-trivial pure state phases, such as long-range order, critical behavior, and topological properties, it is natural to ask the question: Can we extend this concept to include unconventional quantum states unique to mixed quantum states?
The answer is yes.
We have discovered that applying a decoherence to a subsystem of a many-body state, we can realize unconventional mixed-states in the subsystems of undecohered sites, such as mixed SPT and SWSSB states.
This perspective is supported by the fact that taking the ensemble average of such a measured system corresponds to applying maximal strength decoherence to a many-body state.

Interestingly, an example of such a mixed state has already been reported in Ref.~\cite{Lu2026}.  
The authors of Ref.~\cite{Lu2026} argue that for a $G \times S$-protected SPT wavefunction in which $G$ defects carry $S$ charge, local $S$-charge dephasing generically induces SWSSB of $G$ in the resulting mixed state.
As a simplest case, the $Z_2\times Z_2$ cluster state, in which the generators of the subsystem symmetries are $G_1\equiv \prod_{j\in even}X_j$ and $G_2\equiv \prod_{j\in odd}X_j$, exhibits the SWSSB mixed states on the remaining odd subsystem if a local-$X$ decoherence is maximally applied on the even sites.
The findings of Ref.\cite{Lu2026} are consistent with our perspective.

In this work, we demonstrate this perspective using an SPT state protected many subsystem $Z_2$ symmetries. We shall show a sequential generation of non-trivial mixed states by using suitable subsystem decoherences, where the decoherence induces the subsystem charge fluctuations.
From an initial general cluster SPT state with large subsystem symmetries, a sequential charge-decoherence to subsystems induces a series of mixed SPT states with the remaining strong symmetries. Then, the final remaining subsystem exhibits SWSSB. 
That is, we see {\it charge-decoherence reduction hierarchy}. Indeed, we elucidate that in the intermediate decoherence step, an interesting mixed state appears.
On the remaining subsystems the mixed cluster SPT orders exhibit. 
At the stabilizer fixed point, 
this mixed cluster SPT can be described by an equal weight mixture of the cluster SPT affected by projective measurements. 
This mixture is produced by summing possible measurement outcome patterns due to the subsystem decoherences.
At the final step, the final decohered mixed state reaches an equal weight mixture of Greenberger-Horne-Zeilinger (GHZ) states on a remaining single subsystem (at the stabilizer fixed point), which is a prototypical of the SWSSB state~\cite{Ma_PRXQuantum6,Lessa2024_2,wang2026}.

As a result, we show that a hierarchical structure of non-trivial mixed states induced by the sequential subsystem decoherences exists. We also clarify that these non-trivial mixed states can be identified by introducing the R\'{e}nyi-2 correlators~\cite{lee2023,Lessa2024_2,sala2024}. At the stabilizer fixed point, these correlations take the value unity. 
We further investigate the deformation from the stabilizer limits for the appearance mixed SPTs and SWSSB, systematically. We especially show that even in the presence of a perturbation breaking the stabilizer limit and respecting the SPT subsystem symmetries, the appearance mixed SPTs and the final SWSSB state are robust as far as we observe the R\'{e}nyi-2 correlations. This robustness is investigated numerically using a matrix-product-state filtering scheme. 

Our results establish a concrete mechanism by which decoherence can systematically generate a hierarchy of nontrivial mixed-state orders from a single pure SPT phase protected by large on-site symmetries. In contrast to previously studied scenarios where decoherence directly induces SWSSB or average-SPT-like phases, we demonstrate that sequential subsystem charge fluctuations can organize a sequence of intermediate mixed SPTs before reaching the SWSSB state. This reveals a unified framework connecting mixed-state SPT order and SWSSB through decoherence.

The rest of this paper is organized as follows. 
In Sec.~II, we explain general $\alpha$ cluster models and their SPT ground states. 
These states are target states in this work.
In Sec.~III, we discuss our main conjecture.
The general cluster SPT state under sequential subsystem decoherences appear non-trivial mixed SPT states and as a final state, a SWSSB appears. 
We clarify a hierarchical structure for non-trivial mixed states and introduce R\'{e}nyi-2 correlations to characterize these mixed states.
In Sec.~IV, we consolidate our conjecture by employing the stabilizer formalism. We show a concrete example to elucidate how the initial cluster SPT changes under sequential subsystem decoherences. We clearly observe that the mixed states have specific types of stabilizer generators for cluster SPT states. In Sec.~V, we show that our decohered states in stabilizer limit exhibit finite value of the R\'{e}nyi-2 correlators by employing the Choi mapping. In Sec.~VI, we carry out a numerical investigation for the decohered system. In particular, until previous sections, our conjecture is based on the stabilizer limit. Then, a further investigation is considered, apart from the stabilizer limit, how robust the non-trivial mixed states is. 
We numerically examine this issue by using matrix product state simulation based on the Choi mapping.
Section VII is devoted to the conclusion.

\begin{figure*}[t]
\begin{center} 
\vspace{0.5cm}
\includegraphics[width=14cm]{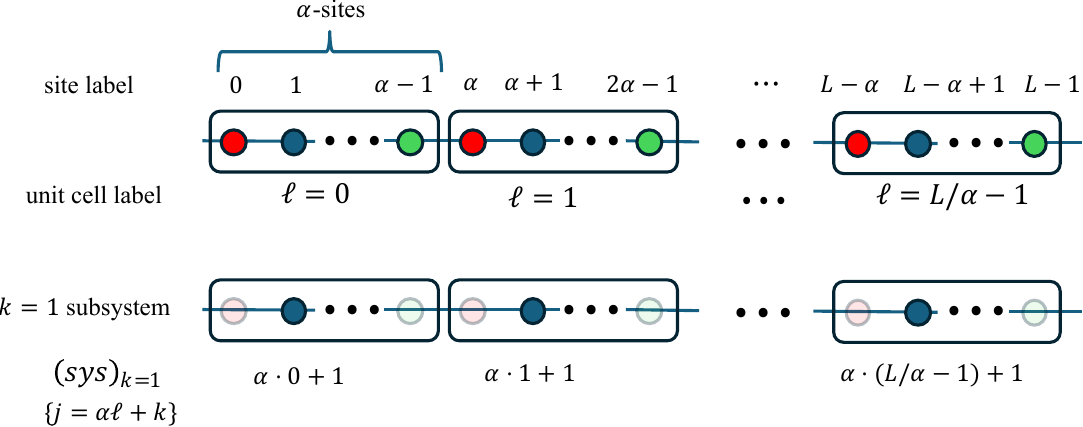}  
\end{center} 
\caption{
Setup of the one-dimensional system with periodic boundary conditions. 
(Upper) The site label is represented by $j=\alpha \ell + k$, where $\ell$ is the unit cell label. The site label $j$ runs $0,1,\cdots, L-1$.
The system includes $\alpha$-different subsystems labeled by $k$ with $k=0,1,\cdots, (\alpha-1)$. The red sites are included in the subsystem $k=0$, the dark blue ones in the subsystem $k=1$, and the green ones in the subsystem $k=\alpha-1$. (Lower) The $k=1$ subsystem is shown. The set of the site on the $k=1$ subsystem is represented by $(sys)_{k=1}$. The site label corresponding to each element of $(sys)_{k=1}$ is given by $\alpha \ell + k$.}
\label{Figlabel}
\end{figure*}
\section{General cluster model and pure SPTs}
In this work, we investigate the effect of local subsystem decoherences on a pure SPT state.
We begin by introducing the following Hamiltonian, namely the general $\alpha$ cluster model \cite{Suzuki1971,Bartlett2009,Smacchia2011,Giampaolo2015,Lahtinen2015}:
\begin{eqnarray}
H_{\rm gc}(\alpha)=-\sum^{L-1}_{j=0}Z_{j}\biggr[\prod^{\alpha-1}_{\ell=1}X_{j+\ell}\biggl]Z_{j+\alpha},
\label{gc_model}
\end{eqnarray}
where $Z_j$, $X_j$ are Pauli operators, $\alpha$ is an integer, and $\alpha \geq 1$ and $j$ (initial) site label, as shown in Fig.~\ref{Figlabel}. Each term $Z_{j}\biggr[\prod^{\alpha-1}_{\ell=1}X_{j+\ell}\biggl]Z_{j+\alpha}$ is a stabilizer \cite{Nielsen_Chuang}, identifying eigenstates of the system by its eigenvalue $\pm 1$. 
Throughout this work, we employ periodic boundary conditions. 

The above cluster model has $\alpha$ subsystem symmetries. The generators are given by \cite{Morral-Yepes2023}: 
\begin{eqnarray}
G^{X,\alpha}_{k}=\prod^{L/\alpha-1}_{\ell=0}X_{\alpha \ell+ k},
\label{symmetry}
\end{eqnarray}
where $k=0, 1, \cdots, \alpha-1$.
For any even $\alpha$, the ground state is the unique gapped SPT state protected by $(Z_2)^{\otimes \alpha}$ symmetry, the set of the generator of which is $\{G^{X,\alpha}_{k}\}$ ($k=0,1,\cdots, \alpha-1$)~\cite{Verresen2017}. 
The most familiar example is $\alpha=2$ case, the ground state of $H_{\rm gc}(2)$ is the cluster state protected by $Z_2\times Z_2$ symmetry\cite{Son2011}. 
For any odd $\alpha$, 
the ground state of $H_{\rm gc}(\alpha)$ is doubly degenerate with spontaneous symmetry breaking (SSB) \cite{Verresen2017} and the cluster SPT state protected by $(Z_2)^{\alpha-1}$ subsystem symmetry, 
where we can pick up $\{G^{X,\alpha}_{k}\}$ with $k=0,1,\cdots,\alpha-2$. The two-fold degeneracy coming from the SSB is distinguished by the sign of the parity operator 
$P\equiv \prod^{L-1}_{j=0} X_j= \prod_{k=0}^{\alpha-1}G_k^{X,\alpha}$, which incorporates $G^{X,\alpha}_{\alpha-1}$. As an example, considering the $\alpha=3$ $(ZXXZ)$ model, the single stabilizer $Z_{j}X_{j+1}X_{j+2}Z_{j+3}$ can be separated as $(Z_{j}Y_{j+1}Z_{j+2})(Z_{j+1}Y_{j+2}Z_{j+3})$. Then, each $(Z_{j}Y_{j+1}Z_{j+2})$ becomes spin-$1/2$ variable and the SSB pattern is implemented on the degree of freedom $(Z_{j}Y_{j+1}Z_{j+2})$ \cite{Verresen2017,Klocke2022}. The ground states of the $ZXXZ$ model are two distinct orthogonal states, each of which corresponds to a cluster SPT state protected by $Z_2\times Z_2$ symmetry~\cite{Verresen2017}. 
In addition, $\alpha=1$ case is just simple. The model is nothing but the Ising model without a transverse field and the ground states are doubly-degenerate $L$-site GHZ states with distinct parity $P=\pm 1$. 

In what follows, we consider the effect of decoherences to the ground state of the Hamiltonian $H_{\rm gc}(\alpha)$. Here, we note that most general $\alpha$ cluster SPT states can be obtained numerically as ground states of the Hamiltonian $H_{\rm gc}(\alpha)$, for example, through imaginary-time evolution starting from a simple product state. Although this provides an efficient numerical method, a more experimentally feasible preparation scheme is desirable. To this end, we present an operational protocol for generating these states from a product state. The details of the protocol are provided in Appendix A.

\section{Sequential subsystem decoherence to a general cluster SPT state}
Based on the stabilizer formalism \cite{Nielsen_Chuang}, we shall give a general discussion that non-trivial mixed states emerge through sequential subsystem decoherences starting from cluster pure SPT state, which is the ground states of $H_{\rm gc}(\alpha)$ for various $\alpha$'s. 
We shall show that the `hierarchical structure' of the non-trivial mixed states appears. To gain a better understanding of the origin of the hierarchical structure, we concretely examine the cases of $\alpha=3$ and $4$ with small system sizes. In these situations, the problem is analytically tractable~\cite{Gottesman1997,aaronson2004}, which allows us to clarify the emergence of the hierarchy in detail.

For the following discussions, we explain the site-setup of the one-dimensional system. The system has $L$ sites, and we assume that $L$ is divisible by $\alpha$.
As shown in Fig.~\ref{Figlabel}, each site is labeled by $j$ as $j=0,1,\cdots L-1$ (site label). Then, we introduce $\alpha$ subsystems. 
Each subsystem includes $L/\alpha=N$ sites. $N$ is the number of unit cells. Each unit cell includes $\alpha$ sites and label by $\ell=0,1,\cdots, N-1$ (``unit cell label'' in Fig.~\ref{Figlabel}). 
Here, we further define the set of sites included in the same subsystem as $(sys)_{k}=\{ j=\alpha \ell +k | \ell=0,\cdots,L/\alpha -1\}$ for $k=0,1,\cdots, \alpha-1$, where $k$ represents the label of subsystems, $\ell$ represents the position of unit cells (unit cell label). $k$ also is regarded as the label of internal sites in each unit cell. 
The schematics for the site labelings and an example subsystem for $(sys)_{k}$ are summarized in Fig.~\ref{Figlabel}.

\subsection{Subsystem decoherences}
We introduce a subsystem decoherence, whose channel representation is given by \cite{Nielsen2011}
\begin{eqnarray}
\mathcal{E}^{X}_{(\alpha,m)}[\rho]&=&\biggl(\prod^{L/\alpha-1}_{\ell=0}\mathcal{E}^{X}_{\alpha \ell +m}\biggl)[\rho],\label{subsystem_channel}
\end{eqnarray}
\begin{eqnarray}
\mathcal{E}^{X}_{\alpha \ell +m}[\rho]&\equiv& (1-p_{X})\rho+p_{X}X_{\alpha \ell +m}\rho X_{\alpha \ell +m},\label{subsystem_channel2}
\end{eqnarray}
where the strength of the decoherence is generally tuned by $p_{X}$, and $0\leq p_{X}\leq 1/2$.
This channel can be expressed by the Lindbladian time-evolution dynamics. 
It is given by $\frac{d\rho}{dt}=\sum_{\ell=0}\frac{1}{2}[X_{\alpha \ell +m}\rho X_{\alpha \ell +m}-\rho]$ with a time interval $t=-\ln(1-2p_X)$. 
The local Lindbladian operation in the sum on the right-hand side can be implemented
in experiments for quantum circuits, as proposed in \cite{PhysRevLett.107.120501,Dong_2023}. 
In this work, we consider step-by-step operation of $\mathcal{E}^{X}_{\alpha \ell +m}$ to an initial state $\rho_0$,
as increasing $m$ through $m=0,1,2,\cdots,\alpha-2$, i.e., $\biggl(\prod^{\alpha-2}_{m=0}\mathcal{E}^{X}_{(\alpha,m)}\biggl)[\rho_0]$.
\begin{figure*}[t]
\begin{center} 
\vspace{0.5cm}
\includegraphics[width=16.5cm]{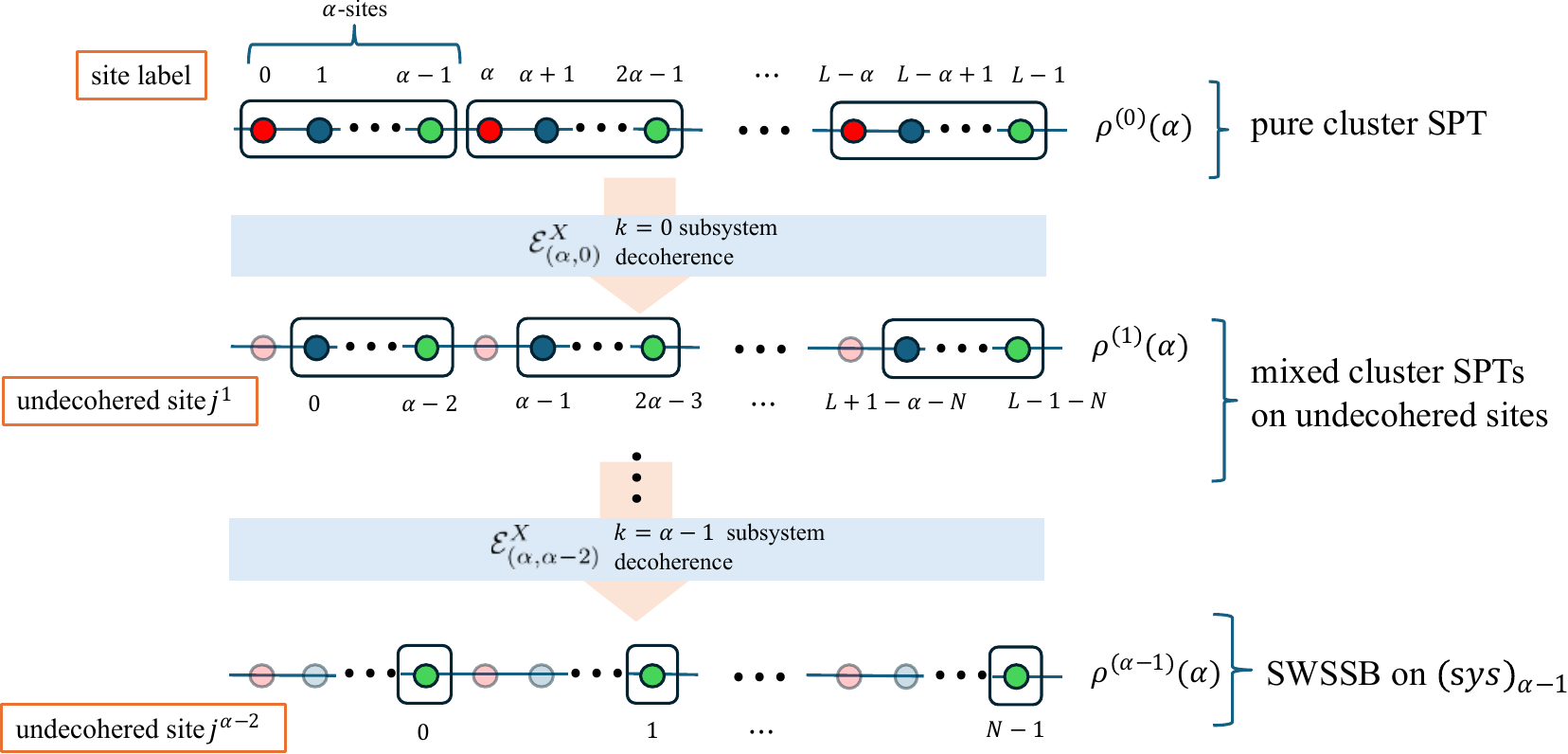}  
\end{center} 
\caption{
Steps of the subsystem decoherence and the obtained mixed states.
The system includes $\alpha$-different subsystems labeled by $k$ with $k=0,1,\cdots, (\alpha-1)$. The red sites are included in the subsystem $k=0$, the dark blue ones in the subsystem $k=1$, and the green ones in the subsystem $k=\alpha-1$. (Upper chain) As the initial setup of the system, we prepare the pure $\alpha$-cluster SPT (pure state). (Middle chain) As the first step of the subsystem decoherence, we apply $\mathcal{E}^{X}_{\alpha,0}$ to the $k=0$ subsystem, $(sys)_{k=0}$ (the faint red sites). On the remaining undecohered sites relabeled by $j^{1}$, a mixed cluster SPT can appear. Such decoherence procedure can be repeated up to the case where the undecohered sites becomes only the $k=\alpha-1$ subsystem, $(sys)_{\alpha-1}$. (Lower chain) The final state after $(\alpha-1)$ times subsystem decoherences can exhibit the SWSSB on the $k=\alpha-1$ subsystem, $(sys)_{\alpha-1}$.}
\label{Fig_deco_system}
\end{figure*}

The operator of the decoherence $X_{\alpha \ell +m}$ corresponds to the charge operator for one of the $Z_2$ symmetries in Eq.~(\ref{symmetry}). Following the expectation proposed in Ref.~\cite{Lu2026}, we conjecture that each $m$-labeled decoherence $\mathcal{E}^{X}_{(\alpha,m)}$ induces a charge-fluctuation effect on the remaining subsystem symmetries. 
In the intermediate stages of the step-by-step decoherences, 
this does not immediately lead to a SWSSB, but instead converts the original pure-state SPT order into lower-rank mixed-state SPT orders protected by the undecohered $Z_2$ subsystem symmetries. 
In this sense, the decohered charge sector acts as a fluctuating background that dresses the defects associated with the remaining symmetries.

Throughout this work, we set $p_X=1/2$, corresponding the maximal decoherence limit. The channel describes projective measurement without recording outcomes. The motivation of the setting comes from the previous studies \cite{Lee2025,lee2025robust} that in one-dimensional system, simple local decoherences tend not to lead a genuinely decoherence-induced phase transition, indeed, there is less possibility to create non-trivial mixed states induced by decoherences.

\subsection{General even-$\alpha$ case} 
We start to discuss a general even-$\alpha$ case. 
The even-$\alpha$ cluster SPT state denoted by $|CS_e(\alpha)\rangle$ is set on $L=\alpha N$ site system. 
Here, we assume the initial state $|CS_e(\alpha)\rangle$ is in a symmetry sector $G^{X,\alpha}_{k}=+1$, that is, the global parity $P=+1$.
The initial (pure) density matrix is $\rho^{(0)}(\alpha)\equiv |CS_e(\alpha)\rangle \langle CS_e(\alpha)|$. 
We perform the decoherence $\mathcal{E}^{X}_{(\alpha,k)}$ on the subsystem sites $(sys)_{k}$ step-by-step. 

As the first step, we apply $\mathcal{E}^{X}_{(\alpha,0)}$ to the state $\rho^{(0)}(\alpha)$. 
As we assume $p_X=1/2$, the decohered state is represented as
\begin{eqnarray}
\rho^{(1)}(\alpha)&=&\sum_{{\vec \beta}^{0}}P^{0}_{{\vec \beta}^{0}} \rho^{(0)}(\alpha)P^{0}_{{\vec \beta}^{0}}.\label{pro1}
\end{eqnarray}
Here, $P^{0}_{{\vec \beta}^{0}}$ is specific for the subsystem decoherence for the general $k$ case as
\begin{eqnarray}
P^{k}_{{\vec \beta}^k}&=&\prod_{j\in (sys)_{k}}\frac{1+\beta_jX_j}{2},
\label{projector}
\end{eqnarray}
where ${\vec{\beta}}^k=\{\beta_{0+k},\beta_{\alpha+k},\cdots, \beta_{\alpha(L/\alpha-1)+k}\}$
is a set of outcomes defined on the subsystem $(sys)_{k}$ corresponding to the eigenvalue of $X_j$ with $\beta_j=\pm 1$. Here, the superscript ``$(1)$'' in the state $\rho^{(1)}(\alpha)$ represents the number of the step of the subsystem decoherence.
Generally, we introduce a label $m$ denoting the number of subsystem decoherence steps, that is, $\rho^{(m)}(\alpha)$ is a decohered state obtained by $m$-step subsystem decoherences,  $\mathcal{E}^{X}_{(\alpha,m-1)}\circ\cdots\circ \mathcal{E}^{X}_{(\alpha,1)}\circ \mathcal{E}^{X}_{(\alpha,0)}\rho^{(0)}(\alpha)=\rho^{(m)}(\alpha)$ with $m\leq \alpha-1$.

We can describe the detailed form of $\rho^{(1)}(\alpha)$. As a first step, we consider how the initial state changes under the projective measurement $P^{0}_{{\vec \beta}^0}$. The initial state changes as follows \cite{Tantivasadakarn2022,verresen2022,Lu2023_feedback,KOI_2024_gc}, 
\begin{eqnarray}
P^{0}_{{\vec \beta}^0}|CS_e(\alpha)\rangle &\propto& |CS^g_e(\alpha-1;{\vec \beta}^{0})\rangle \otimes |{\vec{\beta}}^0_x\rangle_{(sys)_{0}}\nonumber\\
&\equiv& |\Psi^{(1)}_{\vec{\beta}^{0}}\rangle,
\label{1st_md_state}
\end{eqnarray}
where $|CS^g_e(\alpha-1;{\vec \beta}^{0})\rangle$ is a $(\alpha-1)$ cluster SPT state and $|{\vec{\beta}}^0_x\rangle_{(sys)_{0}}$ is a $X$ product state satisfying $X_{j\in (sys)_0}|{\vec{\beta}}^0_x\rangle_{(sys)_{0}}=\beta_j|{\vec{\beta}}^0_x\rangle_{(sys)_{0}}$.  
Here, we note some important points for the obtained state: 
$(\alpha-1)$ cluster SPT state $|CS^g_e(\alpha-1)\rangle$ is in a parity $P_{0}\equiv \prod_{j\in (all)-(sys)_{0}}X_j=1$ where $(all)-(sys)_{0}$ denotes the set of all sites except the measured sites in $(sys)_{0}$ and the allowed outcome configurations are not independent. 
As mentioned in the top of this subsection, since the initial cluster state is chosen in the symmetry sector
$G^{X,\alpha}_0=\prod_{\ell=0}^{N-1}X_{\alpha \ell}=+1$, a nonvanishing measurement branch must satisfy
\begin{eqnarray}
\beta_{\alpha(N-1)}=\prod_{\ell=0}^{N-2}\beta_{\alpha\ell}.
\label{outcome_cond0}
\end{eqnarray}
Furthermore, the total parity decomposes as 
$
P\equiv G^{X,\alpha}_0 P_0
$, the remaining unmeasured subsystem has a parity condition $P_0=+1$.

The state $|CS^g_e(\alpha-1;{\vec \beta}^{0})\rangle$ can be regarded as a ground state of the following effective Hamiltonian given as
\begin{eqnarray}
H^{\rm eff}(1)=
-\sum^{L-N-1}_{j^1=0}\beta_{n^0(j^1)}Z_{j^1}\biggr[\prod^{\alpha-2}_{\ell=1}X_{j^1+\ell}\biggl]Z_{j^1+\alpha-1},\nonumber\\
\label{1st_md_state_effH}
\end{eqnarray}
where the unmeasured sites after the first step measurement are renumbered in order as $j^1$ as shown in Fig.\ref{Figlabel} (a) and the general labeling-rule between the initial site label $j$ and $j^m$ is given in Appendix C. 
The sign of each stabilizer term is determined by the outcomes $\vec{\beta}^0$, where the site label of the outcome $n^{0}[j^1]$ 
denotes the decohered subsystem site in the support of original operator $ZX\cdots XZ$ (stabilizer) to which the site $j^1$ belongs. The labeling-rule is also defined in Appendix C. Then, in terms of the stabilizer formalism \cite{Gottesman1997,Nielsen_Chuang}, the representation of the set of the stabilizer generator for the state $P^{0}_{{\vec \beta}^0}|CS_e(\alpha)\rangle$ is given by $\mathcal{S}^{\alpha-1}=[g^{\alpha-1}_0, g^{\alpha-1}_1, \cdots, g^{\alpha-1}_{L-L/\alpha-1}, P_0]_{(all)-(sys)_{0}}+[\beta_0X_0,\beta_\alpha X_\alpha, \cdots, \beta_{\alpha(L/\alpha-1)}]_{(sys)_{0}}$, where $g^{\alpha-1}_{j^1}=\beta_{n^0(j^1)} Z_{j^1}\biggr[\prod^{\alpha-2}_{\ell=1}X_{j^1+\ell}\biggl]Z_{j^1+\alpha-1}$. 
The projective measurement $P^{0}_{\vec \beta}$ for the initial $\alpha$-cluster SPT state produces the $(\alpha-1)$(odd)-cluster SPT state with $P_0=+1$ appearing on the unmeasured sites. 

Based on the state change by this projective measurement, we can describe the mixed state $\rho^{(1)}(\alpha)$.
Here, we introduce the set of allowed outcomes is given by
\begin{eqnarray}
{\cal B}_{0}
\equiv
\left\{
\vec{\beta}^{0}
\ \middle|\
\beta^{0}_{\alpha \ell}=\pm 1,\quad
\prod_{\ell=0}^{N-1}\beta^{0}_{\alpha \ell}=+1
\right\}.
\end{eqnarray} 
The mixed state generated by the first non-selective subsystem measurement [Eq.~(\ref{pro1})] is then written as the equal-weight convex sum over the allowed branches,
\begin{eqnarray}
&&\rho^{(1)}(\alpha)=
\sum_{\vec{\beta}^{0}\in{\cal B}_{0}}
\frac{1}{|{\cal B}_{0}|}|\Psi^{(1)}_{\vec{\beta}^{0}}\rangle
\langle \Psi^{(1)}_{\vec{\beta}^{0}}|,
\label{rho_1_alpha}
\end{eqnarray}
where $|{\cal B}_{0}|$ is the number of the pattern of the independent outcomes (corresponding to trajectory patterns), $|{\cal B}_{0}|=2^{N-1}$. 
This expression shows that \(\rho^{(1)}(\alpha)\) is a mixed-state cluster SPT.  

From this structure of the state, we directly expect how to characterize this mixed SPT for each fixed outcome $\vec{\beta}^{0}$:  a $(\alpha-1)$-cluster SPT state appears on the unmeasured sites, where the signs of the stabilizer are determined by $\vec{\beta}^{0}$. The non-selective measurement takes an equal-weight mixture over these sign patterns. Thus, the conventional string order of \((\alpha-1)\)-cluster SPT \cite{Morral-Yepes2023} is averaged out. However, we expect that the cluster-SPT structure can be captured by employing the R\'{e}nyi-2 string correlator, discussed later. 

We herein note the aspect of the symmetry. In general, there are two notions of symmetry, weak or strong symmetry for density matrix \cite{Buca_2012,groot2022}, the definitions of which are in Appendix D. For our case, the state $\rho^{(1)}$ keeps the strong symmetry for the initial $(Z_2)^{\otimes \alpha}$ symmetries, that is, $G^{X,\alpha}_{k}\rho^{(1)}(\alpha)=\rho^{(1)}(\alpha)G^{X,\alpha}_{k}=\rho^{(1)}(\alpha)$ with $k=0,\cdots, \alpha-1$. Thus, in this sense, the state $\rho^{(1)}(\alpha)$ is a mixed SPT but is not any averaged SPT since the operator of the decoherence does not reduce the strong $(Z_2)^{\otimes \alpha}$ symmetry of $\{G^{X,\alpha}_{k}\}$ to weak ones \cite{Ma_PRXQuantum6,Lessa2024_2,Lee2025}. 

From the behavior of the single step discussed above, we expect that through the successive decoherence for different $k$ subsystems, this mechanism generates a hierarchy of mixed-state SPT phases, and finally terminates in a SWSSB state with glassy long-range order. Based on this expectation, we next turn to proceed the subsystem decoherences. By employing the subsystem decoherence $\mathcal{E}^{X}_{(\alpha,m)}$ in a sequential manner, we obtain a series of the reduced mixed SPT states with the reduced symmetries defined on the undecohered subsystems. 
We consider the general $m$ ($\leq \alpha-2$) step decohered case. We apply the sequential decoherence $m$-step subsystem decoherences,  $\mathcal{E}^{X}_{(\alpha,m-1)}\circ\cdots\circ \mathcal{E}^{X}_{(\alpha,1)}\circ \mathcal{E}^{X}_{(\alpha,0)}\rho^{(0)}(\alpha)=\rho^{(m)}(\alpha)$, and obtain the mixed state $\rho^{(m)}(\alpha)$. Then, we can describe the components of $\rho^{(m)}$. 
Each component of $\rho^{(m)}(\alpha)$ is indeed the projected states from the initial $\alpha$-cluster SPT obtained by applying the projective measurements $P^{k}_{{\vec \beta}^k}$ with $k=0,1,\cdots, m-1$ to the state $|CS_e(\alpha)\rangle$ with the outcomes $\vec{\beta}^k$. 
Based on the previous studies \cite{Tantivasadakarn2022,verresen2022,Lu2023_feedback,KOI_2024_gc}, we expect that the element has a form,
\begin{eqnarray}
&&\biggr[\prod^{m-1}_{k=0}P^{k}_{{\vec \beta}^k}\biggr]|CS_e(\alpha)\rangle
\propto|CS^g_e(\alpha-m)\rangle \biggr[\bigotimes^{m-1}_{k=0}|\vec{\beta}^k_x\rangle_{(sys)_{k}}\biggl],\nonumber\\
&&\equiv |\Psi^{(m)}_{\vec{\beta}^{0}\cdots \vec{\beta}^{m-1}}\rangle
\label{m_step_element}
\end{eqnarray}
where $|CS^g_e(\alpha-m)\rangle$ can be a cluster SPT state and $|{\vec{\beta}}^k_x\rangle_{(sys)_{k}}$ is a $X$ product state. Under the initial fixing of the parity $P=+1$ and $G^{X,\alpha}_{k}=+1$, the state $|CS^g_e(\alpha-m)\rangle$ can be regarded as the unique ground state of the effective Hamiltonian defined on the remaining unmeasured all sites, given as 
\begin{eqnarray}
&&H^{\rm eff}(m)=\nonumber\\
&&-\sum^{L-mL/\alpha-1}_{j^m=0}
\tilde{\beta}(j^{m})Z_{j^m}\biggr[\prod^{\alpha-1-m}_{\ell=1}X_{j^m+\ell}\biggl]Z_{j^m+\alpha-m}.\nonumber\\
\label{2st_md_state_effH}
\end{eqnarray}
Here $\tilde{\beta}(j^{m})=\prod^{m-1}_{\ell=0}\beta_{q_{\ell}[j^{m}]}$ where $q_\ell[j^{m}]$ is a decohered site between $j^{m}$ and $j^{m}+\alpha-m$, and the unmeasured sites after the $m$-step measurement are labeled in order as $j^m=j^m[j]$, see Appendix C, showing their labeling-rules. 
Here, after $m$-measurements $\prod^{m-1}_{k=0}P^{k}_{{\vec \beta}^k}$, the initial $\alpha$-cluster SPT state is transformed into a $(\alpha-m)$-cluster SPT state $|CS^g_e(\alpha-m)\rangle$ appearing on the unmeasured sites.

From the form of the element, we can straightforwardly describe the $m$-step decohered state  $\rho^{(m)}(\alpha)$ as
\begin{eqnarray}
&&\rho^{(m)}(\alpha)=
\sum_{\vec{\beta}\in{\cal B}_{m-1}}
\frac{1}{|{\cal B}_{m-1}|}|\Psi^{(m)}_{\vec{\beta}^{0}\cdots \vec{\beta}^{m-1}}\rangle
\langle \Psi^{(m)}_{\vec{\beta}^{0}\cdots \vec{\beta}^{m-1}}|,\nonumber\\
\label{rho_m_alpha}
\end{eqnarray}
where $|{\cal B}_{m-1}|$ is the number of elements for the set of allowed outcomes given by
\begin{eqnarray}
{\cal B}_{m-1}
&\equiv&
\{
\vec{\beta}^{k}\ \Big|\ 
\beta^{k}_{\alpha \ell+k}=\pm 1,\quad
\prod_{\ell=0}^{N-1}\beta^{k}_{\alpha \ell+k}=+1,
\nonumber\\
&&\left.
k=0,1,\cdots,m-1
\right\}.
\end{eqnarray}
The mixed state $\rho^{(m)}(\alpha)$ is generated by the non-selective $m$-subsystem measurements as the equal-weight convex sum over the allowed branches.
 
This procedure results in inducing further small symmetric mixed SPT states on the undecohered sites. Finally, after the $(\alpha-1)$-times subsystem decoherences ($m=\alpha-1$), the mixed state $\rho^{(\alpha-1)}(\alpha)$ is expected by considering the projected-measurement element \cite{Tantivasadakarn2022,verresen2022,Lu2023_feedback,KOI_2024_gc}.
Each component is given by
\begin{eqnarray}
&&\biggl(\prod^{\alpha-2}_{k=0} P^{k}_{{\vec \beta}^k}\biggr)|CS_e(\alpha)\rangle\propto|{\rm GHZ}^g_+\rangle_{(sys)_{\alpha-1}} \biggr[\bigotimes^{\alpha-2}_{k=0} |\vec{\beta}^{k}_x\rangle_{(sys)_{k}}\biggl]\nonumber\\
&&\equiv |\Psi^{(\alpha-1)}_{\vec{\beta}^{0}\cdots \vec{\beta}^{\alpha-2}}\rangle.
\label{last_projected_state}
\end{eqnarray}
This state has the $\alpha$-period long-range-ordered state as the `glassy GHZ state' with $P_{\alpha-2}\equiv \prod_{j\in (all)-\sum^{\alpha-2}_{k=0}(sys)_{k}}X_j=1$ denoted by $|{\rm GHZ}^g_+\rangle_{(sys)_{\alpha-1}}$ defined on the subsystem $(sys)_{\alpha-1}$. 
We used the terminology `glassy GHZ state' in the above as the orientation of spin at each unmeasured site varies depending on the outcomes but there still exists long-range entanglement in the resultant subsystem such as 
${1\over \sqrt{2}}(|\uparrow\uparrow\downarrow\cdots\rangle+|\downarrow \downarrow\uparrow\cdots\rangle)$.

From this element $|\Psi^{(\alpha-1)}_{\vec{\beta}^{0}\cdots \vec{\beta}^{\alpha-2}}\rangle$, we can obtain the final mixed state $\rho^{(\alpha-1)}(\alpha)$ as
\begin{eqnarray}
&&\rho^{(\alpha-1)}(\alpha)=
\sum_{\vec{\beta}^{k}\in{\cal B}_{\alpha-2}}
\frac{1}{|{\cal B}_{\alpha-2}|}|\Psi^{(\alpha-1)}_{\vec{\beta}^{0}\cdots \vec{\beta}^{\alpha-2}}\rangle
\langle \Psi^{(\alpha-1)}_{\vec{\beta}^{0}\cdots \vec{\beta}^{\alpha-2}}|,\nonumber\\
\label{rho_final_alpha}
\end{eqnarray}
where $|{\cal B}_{\alpha-2}|$ is the set of allowed outcomes given by
\begin{eqnarray}
{\cal B}_{\alpha-2}
&\equiv&
\{
\vec{\beta}^{k}\ \Big|\ 
\beta^{k}_{\alpha \ell+k}=\pm 1,\quad
\prod_{\ell=0}^{N-1}\beta^{k}_{\alpha \ell+k}=+1,
\nonumber\\
&&\left.
k=0,1,\cdots,\alpha-2
\right\}.
\end{eqnarray}
Here, note that if we focus on the undecohered $k=\alpha-1$ subsystem, there appears the GHZ mixture. This portion just corresponds to a typical SWSSB state \cite{Lu2026}, where the state $\rho^{(\alpha-1)}(\alpha)$ is strongly symmetric to the $Z_2$-symmetry by $G^{X,\alpha}_{\alpha-1}$. The GHZ mixture preserves the existence of a long-range order on the subsystem $(sys)_{\alpha-1}$.

\subsection{General odd-$\alpha$ case} 
Of course, the above discussion satisfies in the case of the initial odd-$\alpha$ cluster SPT state, where we pick up $P=+1$ sector and $G^{X,\alpha}_{k}=+1$ sectors. 
Under the initial sector fixing, the flow of the mixed states under step-by-step subsystem decoherences is basically the same as that of the even-$\alpha$ case. That is, the series of the decohered state coming from the initial cluster SPT exhibits the subsystem mixed SPT states up to the $(\alpha-2)$-step of the subsystem decoherences and the final decohered state $\rho^{(\alpha-1)}(\alpha)$ also exhibits glassy GHZ mixture, that is, has the typical SWSSB mixed part in the undecohered subsystem, $(sys)_{\alpha-1}$. The aspect of the symmetry is also invariant, that is, under the decoherences, the initial strong symmetries protecting the initial SPTs remains and the intermediate mixed SPTs are not ASPT.

\subsection{Characterization for non-trivial mixed states}
The state decohered by the subsystem decoherence has the mixed SPT order, where on each component of the state has the different signs of the stabilizer, depending on the outcome patterns. Indeed, the mixed state can be expressed in the convex sum of the outcome patterns. Then, it is natural to introduce R\'{e}nyi-2 string order for characterizing the mixed-state SPTs.
For $\alpha'$-cluster mixed SPT state in a mixed state $\rho^D$, the R\'{e}nyi-2 string order is introduced as 
\begin{eqnarray}
\mathcal{O}^{st}_{R2}(\alpha',i_0,k_0)[\rho_D]=\frac{\Tr [\rho_D\hat{S}(\alpha',i_0,k_0)\rho_D\hat{S}(\alpha',i_0,k_0)]}{\Tr [\rho^2_D]},\nonumber\\
\label{R2_STO}
\end{eqnarray}
where $i_0$ and $k_0$ are residing sites of charged operators (charged to $X$) corresponding to the edge of the string operator. The form of $\hat{S}(\alpha',i_0,k_0)$ \cite{Morral-Yepes2023} is 
\begin{eqnarray}
\hat{S}(\alpha',i_0,k_0)=Z_{\alpha' i_0}\biggl[\prod^{k_0-1}_{i=i_0}
\biggl(\prod^{\alpha'-1}_{m=1}X_{\alpha' i+m} \biggr)\biggr]Z_{\alpha' k_0}.
\label{STO}
\end{eqnarray}
Here, we note that the supports of the operator $\hat{S}(\alpha',i_0,k_0)$ reside on the undecohered sites.
The undecohered sites are labeled by the relabel function $j^{m}$ after $m$ steps of the subsystem decoherence. 
From the form of the mixed state $\rho^{(m)}(\alpha)$ ($m\leq (\alpha-2)$) in the previous subsection, we expect that the eigenvalue of $\hat{S}(\alpha',i_0,k_0)$ takes $+1$ or $-1$ values randomly reflecting random outcomes $\vec{\beta}$ (under the stabilizer limit). 
However, such a random sign is canceled by the R\'{e}nyi-2 weight $\hat{S}(\alpha',i_0,k_0)$ and then for $\rho^{(m)}(\alpha)$, $\mathcal{O}^{st}_{R2}(\alpha',i_0,k_0)=+1$. This expectation is more concretely observed in Sec.~V.  
On the other hand, we also can consider the conventional string order \cite{Morral-Yepes2023}. We expect that for the mixed SPT state $\rho^{(m)}(\alpha)$, the corresponding conventional string order stays in zero in the stabilizer limit. This expectation is numerically confirmed in Appendix E.

Next, it is natural to introduce 
R\'{e}nyi-2 $ZZ$-correlator
\begin{eqnarray}
\mathcal{O}^{ZZ}_{R2}(i_0,k_0)&=&\frac{\Tr[\rho_D Z_{i_0}Z_{ k_0}\rho_D Z_{i_0}Z_{ k_0}]}{\Tr [\rho^2_D]}.
\label{SGO}
\end{eqnarray}
Here, note that $i_0$, and $k_0$ resides on undecohered sites, i.e.,
we are interested in the correlations in the subsystem $(sys)_{\alpha-1}$. 
It is known that the R\'{e}nyi-2 $ZZ$-correlator is efficient order parameter to identify the SWSSB order \cite{lee2023,sala2024,Ando2026}. 
While as another candidate, a fidelity correlator has been proposed in \cite{Lessa2024_2}.
In this work, we focus on the R\'{e}nyi-2 $ZZ$-correlator since this is numerically more tractable. We expect that the final decohered state $\rho^{(\alpha-1)}(\alpha)$ possesses $\mathcal{O}^{ZZ}_{R2}(i_0,k_0)=+1$ since the final state includes the mixture of the glassy GHZ states.
Then, we also consider the conventional $ZZ$-correlator.
We expect that for the mixed state $\rho^{(\alpha-1)}(\alpha)$, the corresponding conventional $ZZ$-correlator stays in zero in the stabilizer limit. This expectation is numerically confirmed in Appendix E.\\

\section{Concrete example through the stabilizer formalism}
Through the stabilizer update scheme, we further verify the structure of the mixed state under the subsystem decoherence, $\rho^{(m)}(\alpha)$. 
In this section, we observe a small-size system with small $\alpha$ using the update scheme of a set of stabilizer generators shown in \cite{Gottesman1997,aaronson2004,Nielsen_Chuang}.
This concrete verification employing the set of stabilizer generators consolidates that the structure of the state $\rho^{(m)}(\alpha)$ has mixed cluster SPTs and SWSSB order.

We consider $L=\alpha N$ system with $N=3$ and $\alpha=4$.
In the stabilizer formalism, the $\alpha=4$ cluster SPT state is characterized by a set of 12 $ZXXXZ$ stabilizer generators denoted by $\mathcal{S}^{\alpha=4}(m=0)$ ($m$ denotes the number of measurement step), which are given by
$$
\mathcal{S}^{\alpha=4}(m=0)=[g^4_{0},\cdots, g^4_{11}],
$$
where 
$g^4_{j}$ is $j$-th stabilizer generator given by  $g^4_{j}=Z_jX_{j+1}X_{j+2}X_{j+3}Z_{j+4}$ with the initial site label $j$. Then, the state is in the symmetry sector $G^{X,\alpha}_{k}=+1$ ($k=0,1,2,3$), corresponding to the global parity $P=+1$.

We consider the application of $\mathcal{E}^{X}_{(\alpha,0)}$ to the set $\mathcal{S}^{\alpha=4}(m=0)$. Here, since $\mathcal{E}^{X}_{(\alpha,0)}$ is constituted by the multiple of the projective measurements for the $k=0$ subsystem without recording outcomes, 
we focus on the element (a trajectory) of the decohered state $\mathcal{E}^{X}_{(\alpha,0)}[\rho^{(0)}(4)]$ through the set of the stabilizer generator. Thus, we observe the update 
for applying $P^{0}_{{\vec \beta}^0}$ to the set $\mathcal{S}^{\alpha=4}(m=0)$ including the information of outcomes. By employing the measurement update scheme for the set of generators \cite{Gottesman1997,aaronson2004,Nielsen_Chuang}, we obtain the updated set of stabilizer generators, 
\begin{eqnarray}
&&\mathcal{S}^{\alpha=4}(m=0)\xlongrightarrow{P^{0}_{{\vec \beta}^0}}
\mathcal{S}^{\alpha=4}(m=1)[\vec{\beta}_0]\nonumber\\
&&=[\beta_0 X_{0},\beta_4 X_{4},\beta_8    X_{8},\nonumber\\
&&\beta_4g^{3}_1,\beta_4g^{3}_2,\beta_4g^{3}_3,
\beta_8g^{3}_5,\beta_8g^{3}_6,\beta_8g^{3}_7,\nonumber\\
&&\beta_0g^{3}_9,\beta_0g^{3}_{10}
,X_1X_2X_3X_5X_6X_7X_9X_{10}X_{11}],
\label{al=4_m=1}
\end{eqnarray}
where $\beta_0\beta_{4}=\beta_8$, we have used the basic transformation among stabilizer generators several times \cite{Nielsen_Chuang} and re-definitions them such as $g^{3}_{j}=Z_{j^1[j]}XXZ$ and also the last element in the set is the parity $\prod X$, both of which are defined on the unmeasured sites.
The element of the set $\mathcal{S}^{\alpha=4}(m=1)[\vec{\beta}_0]$ gives a straightforward correspondence to the elements of the general form of the state $\rho^{(1)}(\alpha)$ [Eq.~(\ref{rho_1_alpha})], that is, $\mathcal{S}^{\alpha=4}(m=1)[\vec{\beta}_0]$ includes the stabilizer generators of $\alpha=3$ cluster state with different sign depending on the outcomes and the parity operator $P=+1$ preserving the choice of the initial sector and each $\beta_jX_j$ generator indicates the presence of the up or down product $X$ state where its up-down pattern depends on the outcome. 

Thus, if we consider the mixture of the state $\mathcal{S}^{\alpha=4}(m=1)[\vec{\beta}_0]$ in terms of the patterns of $\vec{\beta}_0$, then the mixed state obtained from the set $\mathcal{S}^{\alpha=4}(m=1)[\vec{\beta}_0]$ can be formally represented as
\begin{eqnarray}
&&\rho^{(1)}(4)\longleftrightarrow\sum_{\vec{\beta}_0\in {\cal B}_{0}}\frac{1}{|{\cal B}_{0}|}\mathcal{S}^{\alpha=4}(m=1)[\vec{\beta}_0],
\label{rho_1_alpha_stab}
\end{eqnarray}
Here, the equal weight factor $\frac{1}{|{\cal B}_{0}|}$ can be understood by the fact that the probabilities of an outcome $\beta=\pm 1$ are equal in the stabilizer formalism \cite{Gottesman1997,Nielsen_Chuang}.

As the same with the first step treatment, we can go further subsystem decoherence step. As the second step, we also focus on the elements of trajectory. Indeed, the projective measurement $P^{1}_{{\vec \beta}^1}$ is applied to the set of $\mathcal{S}^{\alpha=4}(m=1)[\vec{\beta}_0]$ and obtain the following set of generators 
\begin{eqnarray}
&&\mathcal{S}^{\alpha=4}(m=1)[\vec{\beta}_0]\xlongrightarrow{P^{1}_{{\vec \beta}^1}}
\mathcal{S}^{\alpha=4}(m=2)[\vec{\beta}_0,\vec{\beta}_1]\nonumber\\
&&=[\beta_0 X_{0},\beta_4 X_{4},\beta_8 X_{8},
\beta_1 X_{1},\beta_5 X_{5},\beta_9 X_{9},\nonumber\\
&&\beta_4\beta_{5}g^{2}_2,\beta_4\beta_{5}g^{2}_3,
\beta_8\beta_{9}g^{2}_6,\beta_8\beta_{9}g^{2}_7,
\beta_0\beta_{1}g^{2}_{10},\beta_0\beta_{1}g^{2}_{11}
],\nonumber\\
\label{al=4_m=2}
\end{eqnarray}
where $g^{2}_j=Z_{j^2[j]}XZ$, and we again find outcomes correlation such as $\beta_9=\beta_1\beta_5$. This set includes the stabilizer generators of the $\alpha=2$ cluster state with a sign pattern depending on the outcomes $\vec{\beta}_{0}$ and $\vec{\beta}_{1}$, defined on the undecohered sites. As in the first step, the equal-weight convex sum over these elements gives the mixed state corresponding to $\rho^{(2)}(4)$ [$m=2$ case of Eq.~(\ref{rho_m_alpha})], which can be formally written as follows
\begin{eqnarray}
&&\rho^{(2)}(4)\longleftrightarrow\sum_{\vec{\beta}_0,\vec{\beta}_1\in {\cal B}_{1}}\frac{1}{|{\cal B}_{1}|}\mathcal{S}^{\alpha=4}(m=2)[\vec{\beta}_0,\vec{\beta}_1].
\label{rho_2_alpha_stab}
\end{eqnarray}

As final step, the element of the set of the final mixed state is obtained by the measurement $P^{2}_{\vec \beta}$ as 
\begin{eqnarray}
&&\mathcal{S}^{\alpha=4}(m=2)[\vec{\beta}_0,\vec{\beta}_1]\xlongrightarrow{P^{2}_{{\vec \beta}^2}}
\mathcal{S}^{\alpha=4}(m=3)[\vec{\beta}_0,\vec{\beta}_1,\vec{\beta}_{2}]\nonumber\\
&&=[\beta_0 X_{0},\beta_4 X_{4},\beta_8 X_{8},
\beta_1 X_{1},\beta_5 X_{5},\beta_9 X_{9},\nonumber\\
&&\beta_2 X_{2},\beta_6 X_{6},\beta_{10} X_{10},\nonumber\\
&&\beta_4\beta_{5}\beta_{6}g^{1}_3, 
\beta_0\beta_{1}\beta_{2}g^{1}_{11},X_3X_7X_{11}
],
\label{al=4_m=3}
\end{eqnarray}
where $g^{1}_j=Z_{j^3[j]}Z$ and $\beta_{10}=\beta_2\beta_6$. 
The stabilizer generators $g^{1}_j$  corresponds to the presence of the glassy GHZ where the domain-wall pattern is determined by the outcomes $\vec{\beta}_0,\vec{\beta}_1$ and $\vec{\beta}_{2}$. The structure of this set corresponds to the form of $\rho^{(\alpha-1)}(\alpha)$ of Eq.~(\ref{last_projected_state}). 
As a result, the final mixed state can be obtained by the equal-weight convex sum over these elements. The mixed state corresponds to $\rho^{(3)}(4)$ [Eq.~(\ref{rho_final_alpha})]. This can be formally written as follows
\begin{eqnarray}
\rho^{(3)}(4)\longleftrightarrow\sum_{\vec{\beta}_0,\vec{\beta}_1,\vec{\beta}_2\in {\cal B}_{2}}\frac{1}{|{\cal B}_{2}|}\mathcal{S}^{\alpha=4}(m=3)[\vec{\beta}_0,\vec{\beta}_1,\vec{\beta}_2].\nonumber\\
\label{rho_3_alpha_stab}
\end{eqnarray}

From this $\alpha=4$ example based on the stabilizer formalism, we verified that the structure of the mixed state for the subsystem decoherences surely has the structure we showed in the previous sections. Indeed, the intermediate mixed state has the mixture of the cluster SPT state on the undecohered sites and at the final step, the last undecohered subsystem are stabilized by $ZZ$ operator with the various signs depending on the outcome, meaning the presence of the glassy GHZ order corresponding to the SWSSB order.

\section{R\'{e}nyi-2 correlators in the stabilizer limit} 
We observed that the decohered states have mixed SPTs and SWSSB order at the stabilizer limit. 
In this section, we show that these phases can be characterized by R\'{e}nyi-2 correlators \cite{lee2023,Lessa2024_2}.
The reader may have already noticed that by inspecting the set of stabilizers, one can indeed show that the R\'{e}nyi-2 correlators take finite values for each decohered states $\rho^{(m)}(\alpha)$, in particular unity.
However, for later convenience, we analytically verify this fact based on an alternative approach, the Choi mapping.

We introduce the Choi mapping \cite{Choi1975,JAMIOLKOWSKI1972}. Firstly, the target Hilbert space $\mathcal{H}$ is doubled as $\mathcal{H}_{u}\otimes \mathcal{H}_{\ell}$, where the subscripts $u$ and $\ell$ denote the upper and lower Hilbert spaces corresponding to bra and ket states of mixed state density matrix, respectively. Then, we can treat the density matrix $\rho$ as a vector, 
$$
\rho \longrightarrow 
|\rho\rangle\rangle\equiv \frac{1}{\sqrt{\dim[\rho]}}\sum_{k}|k\rangle_{u}\otimes \rho|k\rangle_{\ell},
$$ 
where $\{|k\rangle_{u} \}$ and $\{|k\rangle_{\ell} \}$ is an orthonormal set of bases in the Hilbert spaces $\mathcal{H}_{u}$ and $\mathcal{H}_{\ell}$. For example, a pure density matrix is described as $\rho_0 \equiv|\psi_0\rangle \langle \psi_0|\longrightarrow |\rho_0\rangle\rangle\equiv |\psi^*_0\rangle_u|\psi_0\rangle_{\ell}$.

In this formalism, the channel operation of Eqs.~(\ref{subsystem_channel}) and (\ref{subsystem_channel2}) is transformed as
\begin{eqnarray}
\mathcal{E}^{X}_{(\alpha,m)}|\rho\rangle\rangle&=&\biggl(\prod^{L/\alpha-1}_{\ell=0}\mathcal{E}^{X}_{\alpha \ell +m}\biggl)|\rho\rangle\rangle,\label{subsystem_channel_double}
\end{eqnarray}
and each subsystem decoherence acts as
\begin{eqnarray}
&&\mathcal{E}^{X}_{\alpha \ell +m}|\rho\rangle\rangle \nonumber\\
&&\equiv[(1-p_{X})I^u_{\alpha \ell +m}I^\ell_{\alpha \ell +m}+p_{X}X_{\alpha \ell +m}^u X_{\alpha \ell +m}^\ell]|\rho\rangle\rangle, \nonumber\\
\label{subsystem_channel2_double}
\end{eqnarray}
where the labels ``$u$'' and ``$\ell$'' on the operators mean acting on a site residing on the upper(bra) space $\mathcal{H}_{u}$ and lower(ket) space $\mathcal{H}_{\ell}$, and 
$I^{u/\ell}_{\alpha \ell +m}$ represents the identity matrix of upper or lower layers.

Also, in this formalism, the R\'{e}nyi-2 type expectation value can be drastically tractable. The expectation value corresponds to 
$\Tr[\rho\hat{A}^\dagger\rho\hat{B}]/\Tr[\rho^2]=\langle \langle \rho|\hat{A}^{u}\hat{B}^{\ell}|\rho\rangle\rangle/\langle \langle\rho|\rho\rangle\rangle$ \cite{Ma_PRXQuantum6,Lee2025}. The R\'{e}nyi-2 type expectation value can be calculated like an inner product operation in the Choi mapping.

From here, we show the calculation for the R\'{e}nyi-2 correlator for typical cases for our decohered state $\rho^{(m)}(\alpha)$. 
In the Choi mapping, the initial pure state is set as
\begin{eqnarray}
|\rho^{(0)}(\alpha)\rangle\rangle = |CS_e(\alpha)\rangle_u |CS_e(\alpha)\rangle_{\ell}.
\end{eqnarray}
At the stabilizer limit, the R\'{e}nyi-2 string order of $\hat{S}(\alpha,i_0,k_0)$ for the initial state where $i_0$ and $k_0$ are two different sites on $(sys)_{\alpha-1}$ can be calculated. The string operators $\hat{S}(\alpha,i_0,k_0)$ acts as
\begin{eqnarray}
\hat{S}(\alpha,i_0,k_0)_{u}\hat{S}(\alpha,i_0,k_0)_{\ell}|\rho^{(0)}(\alpha)\rangle\rangle=|\rho^{(0)}(\alpha)\rangle\rangle.
\label{Stab_cond}
\end{eqnarray}
Thus, the numerator of the R\'{e}nyi-2 string order
\begin{eqnarray}
&&\langle \langle \rho^{(0)}(\alpha)|\hat{S}(\alpha,i_0,k_0)_{u}\hat{S}(\alpha,i_0,k_0)_{\ell}|\rho^{(0)}(\alpha)\rangle\rangle\nonumber\\
&&=\langle \langle \rho^{(0)}(\alpha)|\rho^{(0)}\rangle\rangle.
\end{eqnarray}
We obtain $\mathcal{O}^{st}_{R2}(\alpha,i_0,k_0)[\rho^{(0)}(\alpha)]=1$.

Next, 
we consider $\rho^{(1)}(\alpha)$ created by the decoherence $\mathcal{E}^{X}_{(\alpha,0)}$, its Choi mapped decohered state is 
\begin{eqnarray}
|\rho^{(1)}(\alpha)\rangle\rangle
=\mathcal{E}^{X}_{(\alpha,0)}|\rho^{(0)}(\alpha)\rangle\rangle
=\sum_{\vec{\beta}_0}P^{0}_{\vec{\beta}^0,u} P^{0}_{\vec{\beta}^0,\ell} |\rho^{(0)}(\alpha)\rangle\rangle\nonumber\\
\end{eqnarray}
Then, there is a relation
\begin{eqnarray}
&&\hat{S}(\alpha-1,i_0,k_0)_{u}\hat{S}(\alpha-1,i_0,k_0)_{\ell}|\rho^{(1)}(\alpha)\rangle\rangle\nonumber\\
&&=\mathcal{E}^{X}_{(\alpha,0)} (\hat{S}(\alpha,i_0,k_0)_{u}\hat{S}(\alpha,i_0,k_0)_{\ell})|\rho^{(0)}(\alpha)\rangle\rangle\nonumber\\
&&=|\rho^{(1)}(\alpha)\rangle\rangle.
\end{eqnarray}
Here, we used $X_{j,u}X_{j,\ell}\mathcal{E}^{X}_{(\alpha,0)}=\mathcal{E}^{X}_{(\alpha,0)}$ with $j\in (sys)_{0}$ and Eq.~(\ref{Stab_cond}). 
Thus, since $\langle\langle \rho^{(1)}(\alpha)|\hat{S}(\alpha-1,i_0,k_0)_{u}\hat{S}(\alpha-1,i_0,k_0)_{\ell}|\rho^{(1)}(\alpha)\rangle\rangle=|||\rho^{(1)}(\alpha)\rangle\rangle||^2$, we obtain $\mathcal{O}^{st}_{R2}(\alpha-1,i_0,k_0)[\rho^{(1)}(\alpha)]=1$. 

By using the same procedure, we straightforwardly obtain ($m\leq \alpha-2$)
\begin{eqnarray}
&&\hat{S}(\alpha-m,i_0,k_0)_{u}\hat{S}(\alpha-m,i_0,k_0)_{\ell}|\rho^{(m)}(\alpha)\rangle\rangle\nonumber\\
&&=
\mathcal{E}^{X}_{(\alpha,m-1)}\cdots \mathcal{E}^{X}_{(\alpha,1)} \mathcal{E}^{X}_{(\alpha,0)}(\hat{S}(\alpha,i_0,k_0)_{u}\hat{S}(\alpha,i_0,k_0)_{\ell})\nonumber\\
&&\times|\rho^{(0)}(\alpha)\rangle\rangle\nonumber\\
&&=\mathcal{E}^{X}_{(\alpha,m-1)}\cdots \mathcal{E}^{X}_{(\alpha,1)}\mathcal{E}^{X}_{(\alpha,0)}|\rho^{(0)}(\alpha)\rangle\rangle.
\end{eqnarray}
Thus, we obtain $\mathcal{O}^{st}_{R2}(\alpha-m,i_0,k_0)[\rho^{(m)}(\alpha)]=1$.

Furthermore, at $m=\alpha-1$ step case, 
the same procedure gives $\mathcal{O}^{ZZ}_{R2}(i_0,k_0)[\rho^{(\alpha-1)}(\alpha)]=1$. 

From these calculations, each decohered states have finite values for each suitable R\'{e}nyi-2 correlators. In the stabilizer limit, these values are unity. This indicates that the mixed states $\rho^{(m)} (\alpha)$ with $m=1,\cdots, \alpha-1$ shown in Sec.III exhibits non-trivial mixed state orders. Here, we give a careful comment that if conventional correlators vanish (see Appendix E) and R\'{e}nyi-2 correlators are finite, then the mixed state is regarded as non-trivial one. 
\begin{figure}[t]
\begin{center} 
\vspace{0.5cm}
\includegraphics[width=6.8cm]{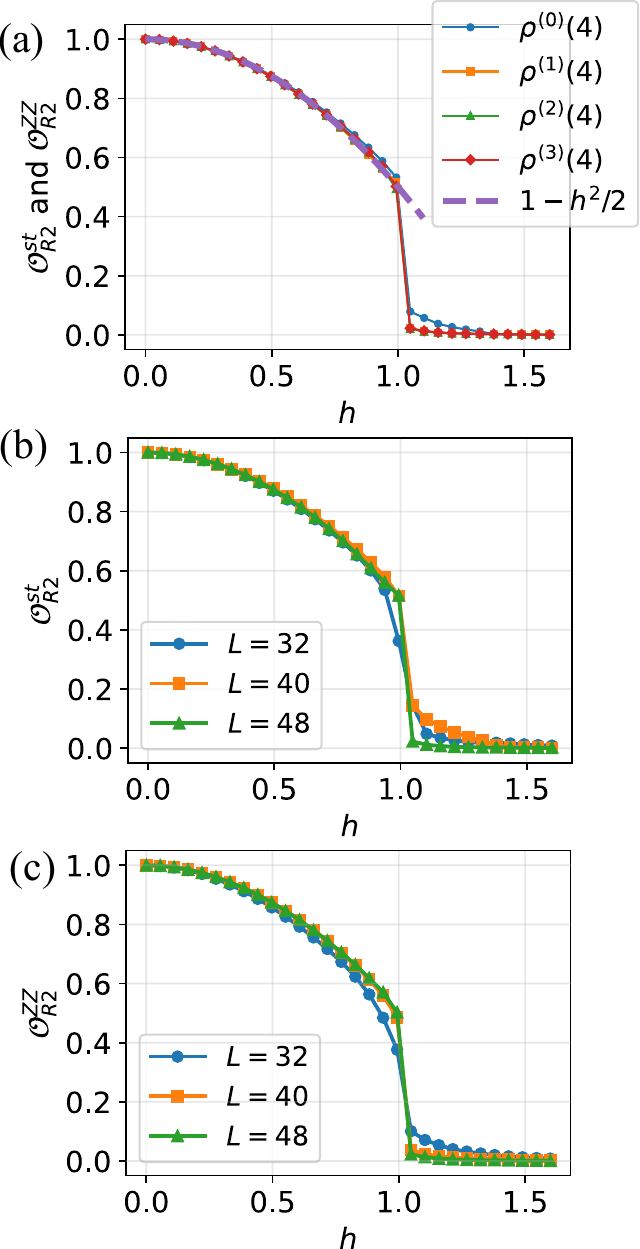}  
\end{center} 
\caption{
(a) R\'{e}nyi-2 string and $ZZ$ correlators for each decohered states $\rho^{(m)}(4)$. 
The purple dashed line is the analytical estimation from the perturbation theory. 
Here, we set $L=48$. 
(b) The system-size dependence of the R\'{e}nyi-2 string correlator with $\alpha=3$ for the decohered state $\rho^{(1)}(4)$.
(c) The system-size dependence of the R\'{e}nyi-2 $ZZ$-correlator for the decohered state $\rho^{(3)}(4)$.
For all data, we set $i_0=\alpha-1$ and $k_0=(\alpha-1)+L/2$ ($L\in \alpha \mathbb{Z}$) with $\alpha=4$. We start to the pure SPT state of $\alpha=4$ cluster model.}
\label{Fig_R2orders_4}
\end{figure}
\section{Numerical calculation --- Mixed state SPTs Beyond Stabilizer Limit}

Taking general $\alpha$ cluster SPT states as concrete examples,
we have given the qualitative discussion and illustrated decoherence-reduction hierarchy based on the stabilizer formalism. 
In what follows, we have numerically shown evidences of the emergent hierarchy structure of mixed state SPTs not only in stabilizer limit, but also in regimes away from stabilizer limit.

To demonstrate that the emergence of hierarchy in mixed SPT states persists away from the stabilizer limit, we consider the general $\alpha$ cluster SPT state under a transverse field magnetic field:
\begin{eqnarray}
H(\alpha)=H_{\rm gc}(\alpha)-h\sum_{j=0}^{L-1} X_j.
\label{gc_h_model}
\end{eqnarray}
The transverse-field terms do not commute with the stabilizer terms in $H_{\rm gc}(\alpha)$, so the ground state of $H(\alpha)$, $|\psi_{GS}\rangle$ is no longer an exact stabilizer state. 
Therefore, the arguments in the previous section are not directly applicable, 
when we use $|\psi_{GS}\rangle$ as an initial state for the subsystem decoherences. 
Nevertheless, since $|\psi_{GS}\rangle$ is in SPT phase for $h<1$, we expect that emergence of hierarchy in mixed state SPTs persists for $h<1$. 
To investigate this expectation, we employ the doubled Hilbert space approach and tensor-network method. 

As introduced in the previous section, the doubled Hilbert space approach, or equivalently Choi mapping allows us to efficiently treat a mixed state density matrix, the decoherence operation and the observables such as R\'{e}nyi-2 correlators. With help of density matrix renormalization method (DMRG), we first prepare the initial density matrix $\rho^{(0)}(\alpha,h)\longrightarrow |\rho^{(0)}(\alpha,h)\rangle\rangle$, where $\rho^{(0)}(\alpha,h)=|\psi_{GS}(\alpha,h)\rangle\langle\psi_{GS}(\alpha,h)|\longrightarrow |\rho^{(0)}(\alpha,h)\rangle\rangle=|\psi_{GS}(\alpha,h)\rangle_u|\psi_{GS}(\alpha,h)\rangle_\ell$, which is equivalent to calculate the ground state of Eq.~(\ref{gc_h_model}) with a decoupled ladder geometry. 
Practically, in numerics, we set the maximum bond dimension $D=200-260$ for ladder system where each upper and lower chain has $L$-sites, the total number of the bond is $3L$, truncate the singular value less than $\mathcal{O}(10^{-8})$, and the energy convergence of the iterative DMRG sweeping is $\Delta E < \mathcal{O}(10^{-6})$ to obtain the initial ground state represented by the matrix product state (MPS).

Then, we apply the subsystem decoherences represented as Eqs.~(\ref{subsystem_channel_double}) and (\ref{subsystem_channel2_double}) to the initial state $|\rho^{(0)}(\alpha,h)\rangle\rangle$ 
\begin{eqnarray}
\mathcal{E}^{X}_{(\alpha,m-1)}\cdots \mathcal{E}^{X}_{(\alpha,1)}\mathcal{E}^{X}_{(\alpha,0)}|\rho^{(0)}(\alpha,h)\rangle\rangle \equiv |\rho^{(m)}(\alpha,h)\rangle\rangle. 
\end{eqnarray}
Using tensor-network method, this operation is efficiently implemented  \cite{Orito2025,KOI2025_v2}.
These numerical procedures can be carried out by employing TeNPy package \cite{TeNPy,10.21468/SciPostPhysCodeb.41-r1.0}.

We now turn to numerical calculations. 
Here, we focus on $\alpha=4$ and $\alpha=3$ cases and observe their mixed states for each decohered steps. For their mixed states, we observe $\mathcal{O}^{st}_{R2}(\alpha-m,i_0,k_0)$ ($m=0,1,\cdots, \alpha-2$) and $\mathcal{O}^{ZZ}_{R2}(i_0,k_0)$ where we fix $i_0=\alpha-1$ and $k_0=(\alpha-1)+L/2$ ($L\in \alpha \mathbb{Z}$).\\

\noindent \underline{$\alpha=4$ result}: 
We show $h$-dependence of $\mathcal{O}^{st}_{R2}(4-m,i_0,k_0)$ ($m=0,1,2$) and $\mathcal{O}^{ZZ}_{R2}(i_0,k_0)$ for each decohered mixed state $\rho^{(m)}(4,h)$. 
For $m=0,1,2$ cases, we observe $\mathcal{O}^{st}_{R2}(\alpha-m,i_0,k_0)$ and for $m=3$, $\mathcal{O}^{ZZ}_{R2}(i_0,k_0)$. The summarized results are shown in Fig.~\ref{Fig_R2orders_4} (a). 
For $h=0$ point (stabilizer limit), the results are consistent to the analysis of the stabilizer limit shown in the previous sections. All correlators have unity. 
For finite $h$ (apart from the stabilizer limit), we find that for $h<1$ regime $\mathcal{O}^{st}_{R2}(4-m,i_0,k_0)$ and $\mathcal{O}^{ZZ}_{R2}(i_0,k_0)$ for each decohered states $\rho^{(m)}(4)$ remains sufficient finite values. These results imply that apart from the stabilizer limit, 
the mixed SPT and the SWSSB orders for the corresponding decohered states $\rho^{(m)}(4,h)$ are robust. Furthermore, the values exhibit sudden drop around $h=1$, implying that $h>1$ the mixed SPT and SWSSB state vanishes. 
Then, we find that $\mathcal{O}^{st}_{R2}(4-m,i_0,k_0)$ and $\mathcal{O}^{ZZ}_{R2}(i_0,k_0)$ for each decohered states $\rho^{(m)}(4)$ takes almost same values although each different $\mathcal{O}^{st}_{R2}(4-m,i_0,k_0)$ and $\mathcal{O}^{ZZ}_{R2}(i_0,k_0)$ cannot be analytically connected. Indeed, from $\mathcal{O}^{st}_{R2}(4-m,i_0,k_0)$ we cannot derive the form of $\mathcal{O}^{st}_{R2}(4-m',i_0,k_0)$ with $m'<m$, which is utterly different situation in the conventional string or ZZ correlations as shown in \cite{Lu2023_feedback,KOI_2024_gc}. Our R\'{e}nyi-2 correlators do not have such inheritance property. However, these values of the R\'{e}nyi-2 correlators have almost same values, same decreasing $h$-dependence. 
We also elucidate that the decreasing curve of the R\'{e}nyi-2 correlators from the unity (stabilizer limit) are $\mathcal{O}(h^2)$-decrease for small $h$ regime. This can be understood by a simple perturbation theory as shown in Appendix F. For small $h$ regime, analytically, R\'{e}nyi-2 correlators behave $1-h^2/2+\mathcal{O}(h^4)$. 
In fact, this estimation is well fitted to the numerical data.
Since the decay is a power-law $h^2/2$, that is, the decrease for small $h$ regime is fairly small, implying that the stabilizer picture are robust.

\begin{figure}[t]
\begin{center} 
\vspace{0.5cm}
\includegraphics[width=6.8cm]{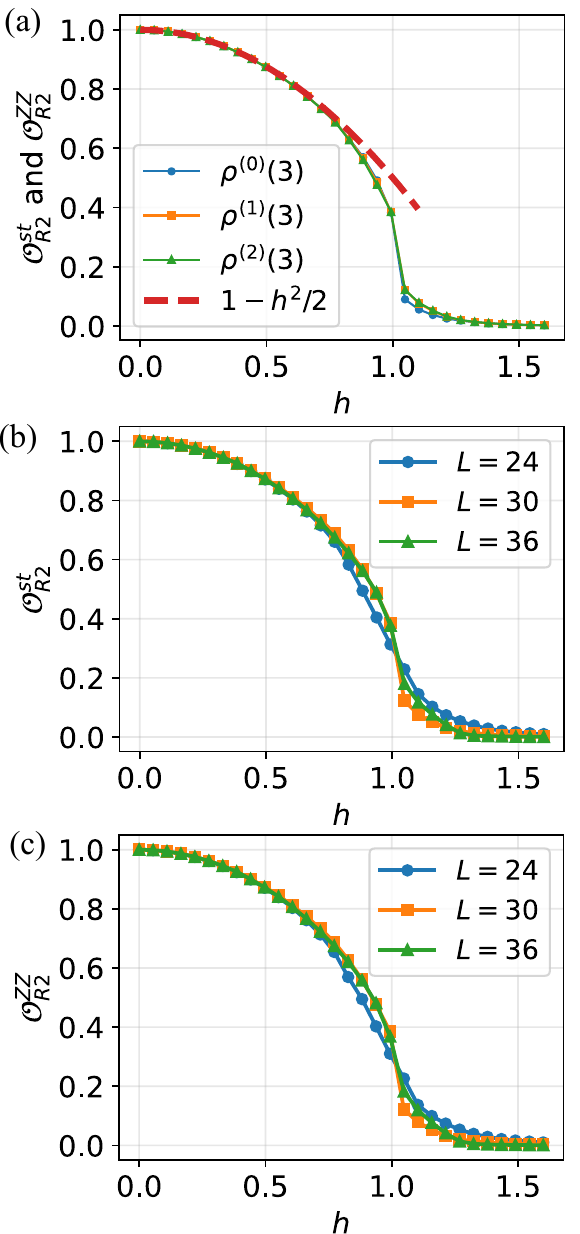}  
\end{center} 
\caption{
(a) R\'{e}nyi-2 string and $ZZ$ correlators for each decohered states $\rho^{(m)}(3)$. The red dashed line is the analytical estimation from the perturbation theory. Here, we set $L=30$. 
(b) The system-size dependence of the R\'{e}nyi-2 string correlator with $\alpha=2$ for the decohered state $\rho^{(1)}(3)$.
(c) The system-size dependence of the R\'{e}nyi-2 $ZZ$-correlator for the decohered state $\rho^{(2)}(3)$.
For all data, we set $i_0=\alpha-1$ and $k_0=(\alpha-1)+L/2$ ($L\in \alpha \mathbb{Z}$) with $\alpha=3$. We start to the pure SPT state of $\alpha=3$ cluster model.}
\label{Fig_R2orders_3}
\end{figure}
We give the further comment on the treatment of $h$-term as a perturbation term. 
In the perturbation theory, the first order perturbed state $|\rho^{(0)'}(4)\rangle\rangle $ (unnormalized initial state) can be written as
\begin{eqnarray}
|\rho^{(0)'}(4)\rangle\rangle = \prod_{\alpha=u,\ell}\biggr[[1+\frac{h}{4}\sum_j X^\alpha_j+\mathcal{O}(h^2)]|CS_e(4)\rangle_\alpha\biggl].\nonumber\\
\end{eqnarray}
Noticeable fact here is that perturbation term, $X_j$ creates a local pair stabilizer excitation; i.e., it flips the eigenvalues of $Z_{j-3}X_{j-2}X_{j-1}Z_{j}$ and $Z_{j}X_{j+1}X_{j+2}Z_{j+3}$ from $1$ to $-1$.
Remarkably, when calculating R\'{e}nyi-2 string order, only local pair stabilizer excitation near the end points of the correlator that anticommute with the end points of $Z$, contribute to the reduction of the R\'{e}nyi-2 string order.
Since the ground state of the $\alpha$-cluster model is expected to be a stabilizer state, based on the perturbation theory, the general $\alpha$-cluster model would also have a similar local pair stabilizer excitation observed in the $\alpha=4$ case~\footnote{For simplicity, we don't take into account the role of logical operator explicitly. While it may play some role, we leave it as a future problem.}.
The perturbative calculation for R\'{e}nyi-2 string order relies solely on commutation and anticommutation of stabilizer generators and the perturbation term $X_j$.
Therefore, we naturally expect that for any $\alpha$ case, the same hierarchical behaviors and $h$ dependence of R\'{e}nyi-2 (string) correlators will appear.

We also show the system-size dependence for the typical decohered states $m=1$ and $3$ as shown in Fig.~\ref{Fig_R2orders_4} (b) and (c). All data indicates that for small $h$ regime, there is no system-size dependence of the R\'{e}nyi-2 correlators. This implies that our practical finite system size sufficiently close to the results for thermodynamic limits and the R\'{e}nyi-2 orders stay in a finite values for thermodynamic limits.\\

\noindent \underline{$\alpha=3$ result}: 
We also examine $\alpha=3$ case. Here, in the initial state preparation by the DMRG, we choose the $P=+1$ sector of the ground state multiplet of the Hamiltonian $H(\alpha)$. Based on the initial $\alpha=3$ pure SPT, we also calculate $h$-dependence of $\mathcal{O}^{st}_{R2}(3-m,i_0,k_0)$ ($m=0,1$) and $\mathcal{O}^{ZZ}_{R2}(i_0,k_0)$ for each decohered mixed state $\rho^{(m)}(3,h)$. 
For $m=0,1$ cases, we observe $\mathcal{O}^{st}_{R2}(\alpha-m,i_0,k_0)$ and for $m=2$, $\mathcal{O}^{ZZ}_{R2}(i_0,k_0)$. The summarized results are shown in Fig.~\ref{Fig_R2orders_3}. 
These results exhibit qualitatively similar behavior to that observed in case $\alpha=4$.
In particular, for small $h$, those correlators exhibit a power-law decay proportional to $h^{2}$, which is in good agreement with perturbative analysis shown in Appendix F.

To further consolidate the presence of non-trivial mixed SPT and SWSSB in our numerics, we verify the vanishing of the conventional string orders and ZZ-correlator for each decohered states $\rho^{(m)}(4,h)$ and $\rho^{(m)}(3,h)$. The results are shown in Appendix E. Surely, the vanishing for these orders keeps in $h<1$ regime. Thus, apart from the stabilizer limit, the non-trivial mixed SPT and SWSSB remains. 

\section{Conclusion}
We have shown a decoherence-induced hierarchy of non-trivial mixed states from the general $\alpha$ cluster SPT states. Based on the stabilizer limit, the sequential subsystem decoherences generate series of the mixed cluster SPT states protected by strong symmetries, which are not average SPTs. 
At the final decoherence step, the decohered mixed state reaches to a SWSSB defined on the undecohered subsystem. 
We clarified that the R\'{e}nyi-2 correlators capture the presence of non-trivial mixed SPTs and SWSSB. Especially, in the stabilizer limit, we showed that the R\'{e}nyi-2 correlators take unity while the conventional string and ZZ correlator vanish.
We further investigate how robust the non-trivial mixed SPTs and SWSSB are apart from the stabilizer limit under a perturbation not breaking the string symmetries. We systematically investigated it by MPS based numerical simulations. Even apart from the stabilizer limit, the R\'{e}nyi-2 correlators remain sufficiently finite. Quantitatively, we found that for small magnetic perturbation, $h^2$ power-law decay appears, implying that the mixed SPTs and SWSSB are robust to the breakdown of the stabilizer picture. 

The present study clarifies how a sequence of subsystem decoherences transforms a single pure SPT phase into distinct non-trivial mixed-states. In particular, we identified a hierarchy of mixed SPT states and its termination into a SWSSB state characterized by glassy GHZ-type long-range entanglement. Our results contribute to a broader understanding of decoherence-induced quantum phases beyond the framework of pure-state quantum matter.

This study provides useful insight into decoherence-based preparation of nontrivial mixed states that have no direct counterparts in pure-state settings. We expect that this proposed scheme, that is, step-by-step decoherence relating to the generator of the system's symmetry may have various possibilities for generating many non-trivial mixed states. 
In particular, SPT or topological order defined on two dimensional system can exhibit various non-trivial mixed states by applying some decoherence relating to subsystem symmetries and 1-form symmetries\cite{mcgreevy2023,Zhang_2025} to their states. Such an issue can be an interesting future work. 
 
\section*{Acknowledgements}
This work is supported by JSPS KAKENHI: 
JP26K06956(Y.K.) and JP26K17056(T.O.). 

\renewcommand{\thesection}{A\arabic{section}} 
\renewcommand{\theequation}{A\arabic{equation}}
\renewcommand{\thefigure}{A\arabic{figure}}
\setcounter{equation}{0}
\setcounter{figure}{0}
\appendix

\setcounter{section}{0}
\renewcommand{\thesection}{\Alph{section}}
\makeatletter
\renewcommand{\theHsection}{\Alph{section}}
\makeatother



\makeatletter
\renewcommand{\theHfigure}{\thesection.\arabic{figure}}
\renewcommand{\theHtable}{\thesection.\arabic{table}}
\renewcommand{\theHequation}{\thesection.\arabic{equation}}
\makeatother
\section{Operation of general cluster SPT states}
We show methods of the state preparation for our target general $\alpha$ cluster SPT states. 
$\alpha=2$ cluster SPT states can be prepared from a simpler state by using the combination of sequential controlled-Z gates (CZ gates), which is sometimes called cluster entangler \cite{Raussendorf2001,Tantivasadakarn2022} and defined by 
\begin{eqnarray}
U_{CZ}\equiv\prod^{L-1}_{j=0}CZ_{j,j+1},
\end{eqnarray}
where $CZ_{j,j+1}$ represents CZ gate for nearest neighbor sites, $j$ and $(j+1)$. 

More generally, general $\alpha$ cluster SPT states can be generated by introducing the pivot transformation~\cite{Tantivasadakarn2023}.
The generator of the pivot transformation denoted by $h^{j}_{k}$ is 
\begin{eqnarray}
U^{p}_{k}=\exp\biggr(i\frac{\pi}{4}\sum_j h^{k}_{j}\biggl),
\end{eqnarray}
where $h^{k}_j$ are given by 
$h^{k}_{j}=Z_jX_{j+1}\cdots X_{j+k-1}Z_{j+k}$ with $k>0$.

In general, this pivot transformation induces the following transformation for $h^{k}_j$ 
\begin{eqnarray}
U^{p}_{k_0}h^{k}_jU^{p\dagger}_{k_0}=h^{2k_0-k}_{j+k-k_0}.
\end{eqnarray}

\noindent \underline{Even $\alpha$ case}:
Now we show how to prepare a general even-$\alpha$ cluster SPT state, $|CS_e(\alpha)\rangle$, which is the \textit{unique ground state} of the Hamiltonian $H_{\rm gc}(\alpha)$. 
First, the $\alpha=2$ cluster SPT state can be created from the $+X$ product state (the unique state) denoted by $|+\rangle^{\otimes L}$ \cite{Raussendorf2001,Tantivasadakarn2022} as
\begin{eqnarray}
|CS_e(2)\rangle=U_{CZ}|+\rangle^{\otimes L}.
\label{state_pre1}
\end{eqnarray}
Based on the state $|CS_e(2)\rangle$, the application of the pivot transformation $U^{p}_{r+2}$ to it creates a general $\alpha=2r+2$ ($r \in \mathbb{N}$) even-$\alpha$ cluster SPT state as 
\begin{equation}
|CS_{e}(\alpha)\rangle=U^{p}_{r+2}|CS_{e}(2)\rangle.
\label{state_pre12}
\end{equation}
This comes from the fact that the Hamiltonian $H_{\rm gc}(2)$ is transformed by the conjugation of the pivot transformation $U^{p}_{r+2}$ as $U^{p}_{r+2}H_{\rm gc}(2)U^{p\dagger}_{r+2} =H_{\rm gc}(2+2r)$.
(Please note that $H_{\rm gc}(2)$ and $U^p_2$ commute with each other.)
By this transformation, the ground state $|CS_e(2)\rangle$ is transformed into $|CS_e(2+2r)\rangle$. 
In this way, we can prepare any even-$\alpha$ cluster SPT state from the simple product state.\\

\noindent \underline{Odd-$\alpha$ case}:
Next, let us discuss the preparation of a general odd-$\alpha$ cluster SPT state denoted by $|CS_{o}(\alpha)\rangle$. 
Note that the ground state is two-fold degenerate in this case \cite{Verresen2017,Morral-Yepes2023}. 
We focus on one of the degenerate ground states, an eigenstate of the parity $P=\prod^{L-1}_{j=0}X_j$.
We start from one of the GHZ ground states of the quantum Ising Hamiltonian $H_{ZZ}=-\sum_{j}Z_jZ_{j+1}$, i.e., the ground state with even parity $P=+1$ such as 
$|{\rm GHZ}_{+}\rangle=\frac{1}{\sqrt{2}}(|\uparrow\rangle^{\otimes L}+|\downarrow\rangle^{\otimes L})$. 
From the state $|{\rm GHZ}_{+}\rangle$, the application of the pivot transformation $U^{p}_{r+1}$ creates a general $\alpha=2r+1$ ($r \in \mathbb{N}$) cluster SPT state as 
\begin{equation}
|CS_{o}(\alpha)\rangle=U^{p}_{r+1}|{\rm GHZ}_{+}\rangle.
\label{state_pre11}
\end{equation}
Here, the state $|CS_{o}(\alpha)\rangle$ keeps the positive parity $P=+1$.

Here, we remark that the pivot transformation for arbitrary $k$ can be implemented by a combination of quantum gates on the quantum circuit. 
By using the combination of the cluster entangler and the pivot transformation, we can prepare any even and odd-$\alpha$ cluster SPT states from the two kinds of states $|+\rangle^{\otimes L}$ and $|{\rm GHZ}_{+}\rangle$, respectively.

\section*{Appendix B: Update scheme of projective measurements in stabilizer formalism}
We explain the efficient update scheme for the projective measurement  \cite{Gottesman1997,aaronson2004,Nielsen_Chuang}. 

The procedure is as follows:\\
Consider a pure state in a $N$-qubit system stabilized by a set of $N$ stabilizer generators.  
Assume that the state is given by the set of the stabilizer generator by $\mathcal{S}=[g_0,g_1,\cdots,g_{N-1}]$, where $g_\ell$ represents an independent stabilizer generator.
For this stabilizer state $\mathcal{S}$, we carry out a projective measurement where the operator of its physical observable $s$ is in a Pauli group $\mathcal{P}_{N}$. From $s^2=1$ the outcome takes $\beta_{s}=\pm 1$. 

Then, by the projective measurement on the state $\mathcal{S}$, the stabilizer generators are updated as follows~\cite{Gottesman1997,aaronson2004}:
\begin{itemize}
\item[(I)] Search anticommutative stabilizer generators to $s$. 
This can be carried out by using the check matrix \cite{Nielsen_Chuang}. 
From this procedure, as the case 1, we obtain single or some $m$ anticommutative stabilizer generators, $g_{\ell_1},g_{\ell_2},\cdots,g_{\ell_m}$ ($m\leq N$). As the case 2, there is no anticommutative one, $\mathcal{S}$ is not updated.

\item[(II)] If the case 1 occurs in (I) and there is only a single stabilizer generator denoted by $g_{m_1}$ anticommute to $s$, we replace $g_{m_1}$ with $\beta_s s$. $\beta_s$ is the outcome with probability $p_{\beta_s}=\langle \Psi_{st}|P_{\beta_s}|\Psi_{st}\rangle=1/2$ with $\displaystyle{P_{\beta_s}=\frac{1}{2}[I+\beta_s s]}$\cite{Nielsen_Chuang}, where $|\Psi_{st}\rangle$ is the stabilizer state by $\mathcal{S}$. The update of $\mathcal{S}$ is achieved. 

\item[(III)] When there are (more than two) $m$ anticommutative stabilizer generators $g_{\ell_1},g_{\ell_2},\cdots, g_{\ell_m}$ ($m\leq N$), replace $g_{\ell_1}$ with $\beta_s s$. Furthermore, for the rest of anticommutative ones $g_{\ell_i}$, update $g_{\ell_i} \to g_{\ell_i} g_{\ell_1}$. By this procedure the set of stabilizer generators $\mathcal{S}$ is updated by the projective measurements with the measurement operator $s$.
\end{itemize}
When we carry out a next projective measurement with a measurement operator $s'$ with the outcome $\beta_{s'}$, we do the above update prescription again with respect to the previous outcome factor $\beta_s$, such as $\beta_s s$.


\section*{Appendix C: Relabel functions}
In this appendix, we show the re-labeling rule for the undecohered sites after $m$ ($\leq \alpha-1$) steps of the subsystem decoherence.
First, $j$ is the initial site label as shown in Fig.~\ref{Figlabel}. 
Then, the site-relabeling after the first ($m=1$) step of the subsystem decoherence,
\begin{eqnarray}
j^1[j]\equiv(\lfloor \frac{j}{\alpha} \rfloor)(\alpha-1)+ [(j\bmod \alpha)-1], \nonumber
\end{eqnarray}
for $j \in (all)-(sys)_0$. 
The site label $j^1$ labels correctly the undecohered sites in order as shown in Fig.~\ref{Fig_deco_system}. 

Generally, the site-relabeling after $m$ steps of the decoherence denoted as $j^{m}$ is given by
\begin{eqnarray}
j^{m}[j]\equiv(\lfloor \frac{j}{\alpha} \rfloor)(\alpha-m)+ [(j\bmod \alpha)-m], \nonumber
\end{eqnarray}
for $j \in (all)-\sum^{m-1}_{k=0}(sys)_k$.

Also, in the effective Hamiltonian after the first step of the decoherence, the site-label of the outcome factor $\beta$ is given by  
\begin{eqnarray}
n^{0}[j^1]\equiv (\lfloor \frac{j^1}{\alpha-1} \rfloor+1)\alpha+0. \nonumber
\end{eqnarray}


\section*{Appendix D: Weak and strong symmetries}
For density matrix, two symmetry notions can be introduced \cite{Buca2012,groot2022}. Accordingly, 
for quantum channels the two notions is available. 

In this work, we focus on the $(Z_2)^{\otimes \alpha}$ symmetries these generators of which is $\{G^{X,\alpha}_k\}$ introduced in the main text. 
Each $Z_2$ symmetry group the group is denoted as $\{\hat{1},G^{X,\alpha}_k\}$. 

First, we give the general definition of the strong symmetry expressed in terms of density matrix,
$$
G^{X,\alpha}_k\rho=e^{i\theta}\rho,\;\;\; \rho G^{X,\alpha\dagger}_k=e^{-i\theta}\rho,
$$
where $\rho$ is a density matrix and $\theta$ is a global phase factor. 

Next the weak symmetry can be defined. Its operation is a conjugation as
$$
G^{X,\alpha}_k\rho G^{X,\alpha\dagger}_k = \rho.
$$ 
This symmetry is in ensemble average level \cite{ma2024}.

The notion of the strong and weak symmetries are introduced on quantum channels. 
Generally, consider the Kraus-operator-sum 
form \cite{Nielsen2011,lidar2020} for the quantum channel
$$
\mathcal{E}(\rho)\equiv\sum^{N-1}_{\ell=0}K_{\ell} \rho K^\dagger_{\ell},
$$ where $\{K_{\ell}\}$ are a set of Kraus operators satisfying $\sum^{N-1}_{\ell=0} K^\dagger_{\ell} K_{\ell}=\hat{I}$ with $\hat{I}$ being the identity 
operator, preserving $\mathrm{Tr}[\mathcal{E}(\rho)]=1$. 
Then, we can apply the Kraus operators to two notion of symmetry strong symmetry. 
The strong symmetry satisfies if 
$$
K_{\ell}G^{X,\alpha\dagger}_k=e^{i\theta} G^{X,\alpha}_k K_{\ell}
$$ 
for any $\ell$ and arbitrary $\theta$. 
On the other hand, the weak symmetry satisfies if 
$$
G^{X,\alpha}_k\biggl[\sum_{\ell}K_{\ell} \rho K^\dagger_{\ell}\biggr]G^{X,\alpha\dagger}_k=\mathcal{E}(\rho).
$$
This definition does not require that each Kraus operator commutes with the generator $G^{X,\alpha}_k$. 
If the strong symmetry is satisfied, the weak one is satisfied. 
In a SWSSB, the density matrix keeps the strong symmetry condition. Thus, the weak symmetry is automatically satisfied.

\begin{figure}[t]
\begin{center} 
\vspace{0.5cm}
\includegraphics[width=6cm]{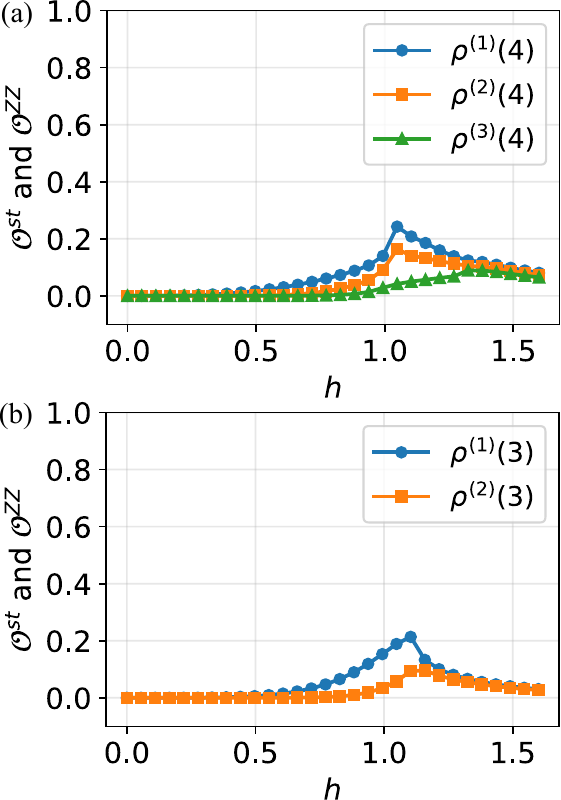}  
\end{center} 
\caption{
(a) Conventional string orders and $ZZ$-correlator for each decohered states $\rho^{(m)}(4)$ from $\alpha=4$ initial cluster SPT state. Here, we set $L=32$. 
(b) Conventional string order and $ZZ$-correlator for each decohered states $\rho^{(m)}(3)$ in $\alpha=3$ initial cluster SPT. Here, we set $L=30$.
For all data, we set $i_0=\alpha-1$ and $k_0=(\alpha-1)+L/2$ ($L\in \alpha \mathbb{Z}$).}
\label{fig:conventional_order}
\end{figure}
\section*{Appendix E: Conventional correlators}

In this Appendix, we show numerical results of the conventional correlation functions. 
One of them is the conventional string order 
\begin{eqnarray}
\mathcal{O}^{st}(\alpha,i_0,k_0)[\rho]=\Tr[\rho\hat{S}(\alpha,i_0,k_0)].
\label{R1_STO}
\end{eqnarray}
In the doubled Hilbert space formalism, the form of $\mathcal{O}^{st}(\alpha,i_0,k_0)[\rho]$ is given by 
\begin{eqnarray}
\mathcal{O}^{st}(\alpha,i_0,k_0)[\rho]=\frac{\langle\langle {\rm ME}|\hat{S}(\alpha,i_0,k_0)|\rho\rangle\rangle}{\langle\langle {\bf 1}|\rho\rangle\rangle},
\end{eqnarray}
where $|{\rm ME}\rangle\rangle\equiv \displaystyle{\frac{1}{2^{3L/2}}\prod^{L-1}_{j=0}|t\rangle_j}$ with
$|t\rangle_j=|\uparrow_u\uparrow_{\ell}\rangle_j+|\downarrow_u\downarrow_{\ell}\rangle_j$. This state $|{\rm ME}\rangle\rangle$ corresponds to the maximally entangled rung-pairs. $\mathcal{O}^{st}(\alpha,i_0,k_0)[\rho]$ is a conventional string order parameter characterizing $\alpha$ cluster SPT \cite{Morral-Yepes2023}.

The second one is ZZ-conventional correlator given by
\begin{eqnarray}
\mathcal{O}^{ZZ}(i_0,k_0)[\rho]&=&\Tr[\rho Z_{i_0}Z_{ k_0}].
\label{SGO_R1}
\end{eqnarray}
In the doubled Hilbert space formalism, the form of $\mathcal{O}^{ZZ}_{R1}(i_0,k_0)[\rho]$ is given by 
\begin{eqnarray}
\mathcal{O}^{ZZ}(i_0,k_0)[\rho]=\frac{\langle\langle {\rm ME}|Z_{i_0,u}Z_{k_0,u}|\rho\rangle\rangle}{\langle\langle {\bf 1}|\rho\rangle\rangle}.
\end{eqnarray}
$\mathcal{O}^{ZZ}(i_0,k_0)[\rho]$ is a conventional order parameter characterizing the conventional SSB (i.e., weak symmetry SSB) \cite{lee2023,Lessa2024_2,sala2024}.

As discussed in the main text, if the decohered states appear the mixed state SPTs or the SWSSB, the conventional correlator vanishes since the pattern of measurement outcomes in the mixed states cancels finite contribution for these correlators. We numerically verify this behavior for the $\alpha=4$ and $\alpha=3$ cases where we employ the same setting to that of the main text.

Figure~\ref{fig:conventional_order} shows the conventional string orders and the conventional $ZZ$ correlator for the decohered states in the $\alpha=4$ and $\alpha=3$ cases.
For both cases, these conventional correlators are strongly suppressed in the $h<1$ region, where the decohered states are expected to realize mixed-state SPT orders.
This behavior is consistent with the fact that the outcome averaging destroys conventional charged correlations, even though the mixed SPT order remains detectable by the R\'{e}nyi-2 string order.
Around the transition region, the correlators show small peaks, indicating enhanced fluctuations near the phase boundary while these peaks do not develop in the mixed SPT and SWSSB regime.
Thus, the numerical results support the interpretation that the decohered states are not characterized by ordinary pure SPT string and order weak-symmetry SSB.

\widetext
\section*{Appendix F: Perturbation theory apart from the stabilizer limit}
We show a perturbation theory (first order) where we treat $h$-term in $H(\alpha)$ of Eq.(\ref{gc_h_model}) as a perturbation term and observe how the R\'{e}nyi-2 string order deviates from the behavior of stabilizer limit for small $h$-term. 

Here, we consider $\alpha=4$ cluster SPT system and observe the R\'{e}nyi-2 string order for the mixed state $\rho^{(1)}(4,h)$ obtained from the initial $\alpha=4$ cluster SPT $\rho^{(0)}(4,h)$, which is not a stabilizer state. From the perturbation theory, the first order perturbed state for $\rho^{(0)}(4,h)$ is given by the following Choi mapped vector
\begin{eqnarray}
|\rho^{(0)'}(4)\rangle\rangle = \biggr[[1+\frac{h}{4}\sum_j X^u_j]|CS_e(4)\rangle_u\biggl] \biggr[[1+\frac{h}{4}\sum_j X^\ell_j]|CS_e(4)\rangle_{\ell}\biggl].
\label{first_pertuerbed_state}
\end{eqnarray}
From the state $|\rho^{(0)'}(\alpha)\rangle\rangle$, the first step decohered state is given by $\mathcal{E}^{X}_{4,0}|\rho^{(0)'}(\alpha)\rangle\rangle$. For this decohered state, the part of the $\alpha=3$ R\'{e}nyi-2 string order is
\begin{eqnarray}
&&\langle\langle \rho^{(0)'}(4)|\mathcal{E}^{X\dagger}_{4,0}\hat{S}(3,i_0,k_0)_{u}\hat{S}(3,i_0,k_0)_{\ell}\mathcal{E}^{X}_{4,0}|\rho^{(0)'}(4)\rangle\rangle\nonumber\\
&&=\langle\langle \rho^{(0)'}(4)|\mathcal{E}^{X\dagger}_{4,0}\mathcal{E}^{X}_{4,0}\biggr(\hat{S}(4,i_0,k_0)_{u}\hat{S}(4,i_0,k_0)_{\ell}|\rho^{(0)'}(4)\rangle\rangle\biggl).
\label{R2_m=1_corr}
\end{eqnarray}
Here by substituting the perturbed state of Eq.~(\ref{first_pertuerbed_state}), the parentheses in the above further become 
\begin{eqnarray}
\biggr(\hat{S}(4,i_0,k_0)_{u}\hat{S}(4,i_0,k_0)_{\ell}|\rho^{(0)'}(4)\rangle\rangle\biggl)&=&|\rho^{(0)}(4)\rangle\rangle+\frac{h}{4}|CS_e(4)\rangle_u\biggl(\hat{S}(4,i_0,k_0)_{\ell}(\sum_jX_j)|CS_e(4)\rangle_\ell\biggr)\nonumber\\
&+&\frac{h}{4}\biggl(\hat{S}(4,i_0,k_0)_{u}(\sum_jX^u_j)|CS_e(4)\rangle_u\biggr)|CS_e(4)\rangle_\ell+\mathcal{O}(h^2).
\end{eqnarray}

By using the above relation and carrying out some calculations, we obtain the RHS of equation (\ref{R2_m=1_corr}) as 
\begin{eqnarray}
&&\langle\langle \rho^{(0)'}(4)|\mathcal{E}^{X\dagger}_{4,0}\hat{S}(3,i_0,k_0)_{u}\hat{S}(3,i_0,k_0)_{\ell}\mathcal{E}^{X}_{4,0}|\rho^{(0)'}(4)\rangle\rangle
\sim\biggr(1+\frac{h^2}{8}(L-4)\biggl)\langle\langle \rho^{(0)}(4)|\mathcal{E}^{X\dagger}_{4,0}\mathcal{E}^{X}_{4,0}|\rho^{(0)}(4)\rangle\rangle.
\label{R2_m=1_corr_nume}
\end{eqnarray}
Then, by employing the same calculation for the norm $\langle\langle \rho^{(0)'}(4)|\mathcal{E}^{X\dagger}_{4,0}\hat{S}(3,i_0,k_0)_{u}\hat{S}(3,i_0,k_0)_{\ell}\mathcal{E}^{X}_{4,0}|\rho^{(0)'}(4)\rangle\rangle$, we obtain the perturbed $\alpha=3$ R\'{e}nyi-2 string order
\begin{eqnarray}
\mathcal{O}^{st}_{R2}(3,i_0,k_0)[\rho^{(1)'}(4)]&=&\frac{\langle\langle \rho^{(0)'}(4)|\mathcal{E}^{X\dagger}_{4,0}\hat{S}(3,i_0,k_0)_{u}\hat{S}(3,i_0,k_0)_{\ell}\mathcal{E}^{X}_{4,0}|\rho^{(0)'}(4)\rangle\rangle}{\langle\langle \rho^{(0)'}(4)|\mathcal{E}^{X\dagger}_{4,0}\mathcal{E}^{X}_{4,0}|\rho^{(0)'}(4)\rangle\rangle}\nonumber\\
&\sim& \frac{\biggr(1+\frac{h^2}{8}(L-4)\biggl)\langle\langle \rho^{(0)}(4)|\mathcal{E}^{X\dagger}_{4,0}\mathcal{E}^{X}_{4,0}|\rho^{(0)}(4)\rangle\rangle}{\biggr(1+\frac{h^2}{8}L\biggl)\langle\langle \rho^{(0)}(4)|\mathcal{E}^{X\dagger}_{4,0}\mathcal{E}^{X}_{4,0}|\rho^{(0)}(4)\rangle\rangle}=1-\frac{h^2}{2}+\mathcal{O}(h^4).
\label{R2_m=1_corr_nume}
\end{eqnarray}
As a result, we evaluated that the R\'{e}nyi-2 string order for the state $\rho^{(1)'}(4)$ exhibits $h^2$ power-law decrease for the small $h$ regime. 
Thus, the decay is not linear but 2nd order power law in terms of $h$. In this sense, the picture of the stabilizer limit sufficiently remains for small $h$ case. 
This concrete calculation for the $\alpha=4$ and $m=1$ case we showed also is applicable to other $\alpha$ and $m$ cases. Similarly, for the corresponding R\'{e}nyi-2 correlators, the same power-law decay in terms of $h$ exhibit. 

Numerical data shown in Figs.~\ref{Fig_R2orders_4} and ~\ref{Fig_R2orders_3} in the main text surely follows this power-law decay for small $h$ regime ($h \lessapprox	1$).

\endwidetext

\nocite{apsrev42Control}
\bibliographystyle{apsrev4-2}
\bibliography{ref}
\end{document}